\documentclass{article}
\usepackage{amsmath}
\usepackage{amssymb}

\input{tcilatex}
\begin{document}

\author{J. G. Cardoso \\
Department of Mathematics\\
Centre for Technological Sciences-UDESC\\
Joinville 89223-100, Santa Catarina,\\
Brazil.\\
e-mail: dma2jgc@joinville.udesc.br
KEY\ WORDS:\\
Relativistic quantum mechanics, Cartan's spaces,\\
Pseudo-Hermitian operators, Orthocronous proper Poincar\'{e} \\
group, Observational correlations, Free twofold systems.}
\title{Pseudo-Unitary Dynamics of Free Relativistic Quantum Mechanical
Twofold Systems}
\date{ \ }
\maketitle

\begin{abstract}
A finite-dimensional pseudo-unitary framework is set up for describing the
dynamics of free elementary particles in a purely relativistic quantum
mechanical way. States of any individual particles or antiparticles are
defined as suitably normalized vectors belonging to the
two-complex-dimensional spaces that occur in local orthogonal decompositions
of isomorphic copies of Cartan's space. The corresponding dynamical
variables thus show up as bounded pseudo-Hermitian operator restrictions
that possess real discrete spectra. Any measurement processes have to be
performed locally in orthocronous proper Lorentz frames, but typical
observational correlations are expressed in terms of symbolic configurations
which come from the covariant action on spaces of state vectors of the
Poincar\'{e} subgroup of an adequate realization of $SU(2,2)$. The overall
approach turns out to supply a supposedly natural description of the
dynamics of free twofold systems in flat spacetime. One of the main outlooks
devised here brings forward the possibility of carrying out methodically the
construction of a background to a new relativistic theory of quantum
information.
\end{abstract}

\section{Introduction}

In both of the traditional pictures of non-relativistic quantum mechanics,
the states of any dynamical systems usually come into play as normalized
elements of complex separable Hilbert spaces which are not at all associated
with unitary representations of the Galilei group [1-3]. Therefore, the
observational correlations that can be sorted out on the basis of the inner
structure of such quantum mechanical contexts just arise from Pauli's $SU(2)$%
-spin theory [4, 5]. Accordingly, any preparations and measurements of spin
one-half states that are eventually performed locally by an observer can be
systematically manipulated by another local observer by carrying out some $%
SO(3)$-transformation. Consequently, the only correlations between
measurement outcomes that may be put into practice within the standard
non-relativistic quantum mechanical framework, bear strictly a combination
of locality with a spin character.

The overall formulation of non-relativistic quantum theory predicted the
well-established experimental fact that unbound spin one-half elementary
particles should be looked upon as twofold systems. This remarkable
dynamical feature had already been widely spread in connection with the
classical electromagnetic description of photon polarizations [6]. It was
made even more transparent with the advent of Dirac's relativistic theory of
electrons and positrons [7] in which all the pertinent charge and spin
degrees of freedom are automatically taken up by the theoretical scope from
the beginning. The values of the total energy of a free Dirac particle are
frequently picked up for some local purposes by describing the dynamics in
the rest frame of the particle. Such a procedure gives rise to a twofold
energy spectrum, and apparently yields the occurrence of a loss of
covariance which is related to a kinematical indetermination\ of the
helicity of the particle. Any pairs of free particle-antiparticle companions
thus generally carry total energies of opposite signs while a recovery of
helicities is accomplished in each case by performing suitable Lorentz
transformations. The Pauli-Dirac twofold features have been carried over as
physical two-valued properties of other degrees of freedom to all of the
major particle schemes [8] brought forward after the presentation of Dirac's
theory. Among these, of course, is the standard description of massless
fermions [9] which notably exhibits a characteristic anomaly associated to
helicity degeneracies.

Nonetheless, from a purely quantum mechanical point of view, that is to say,
without effectively regarding any structural aspects of the existing quantum
field theories, the theory of elementary particles as it stands at the
present time, bears a flawful character in that no conceivable fundamental
decompositions involving simultaneously spaces of state vectors for free
particles and antiparticles really emerge thereabout. Noticeably enough,
such an imperfection takes place even when one calls for the faithful
representation of the orthocronous proper component $\mathcal{L}%
_{+}^{\uparrow }$ of the Lorentz group along with the $SL(2,\mathbf{C})$%
-spinor version of Dirac's theory as designed originally by van der Waerden
[10]. Within this framework, any dynamical state appears as a non-orthogonal
direct sum between a pair of two-complex-component spinors that always
describe covariantly the admissible helicities of one and the same particle,
there being likewise a locally defined space of states for the particle at
issue which is endowed with a definite inner product. The entire two-level
description is then brought out when some total-energy and electric-charge
values are appropriately ascribed to certain conjugate states. It becomes
evident that the conventional relativistic quantum mechanical scenario
affords local observational correlations in Minkowski space without making
it feasible to cope with unitarity in any fundamental way.

The situation concerning the lack of intrinsic unitarity in Dirac's theory
was circumvented by the construction of the famous Wigner classification
schemes for elementary particles [11-14]. Roughly speaking, such schemes
include building up explicit irreducible representations of the orthocronous
proper Poincar\'{e} group $\mathcal{P}_{+}^{\uparrow }$ towards achieving a
unitary description of the spacetime behaviours of wave functions for free
particles wherein all spectral contributions coming from orbital
angular-momentum generators are invariantly equal to zero. It appears that
the representations for any spin-$s$ massive particles carrying either
positive or negative energies, bear $(2s+1)$ discrete labels. In the case of
both energy characters, these representations are completely specified by
assembling the relevant spin labels and the linear-momentum components of
time-like energy-momentum four vectors. The representations for spinning
massless particles of either energy type, on the other hand, admit\ both
discrete and continuous spin labels, but only pairs of discrete helicity
degrees of freedom bear physical meaningfulness. For given null
energy-momentum four vectors, any representations of this latter kind
provide dynamical descriptions which carry pairs of spin labels. One then
becomes able to write down the observational correlations for the spectral
configurations of any relativistic theories as similarity transformations
that involve unitary operators acting on the respective representation
spaces. A definite inner product is set upon any such representation space
which, therefore, gets identified with a Hilbert space. However, any
representations for particles come about apart from any others for
antiparticles whence no representation is produced which fits together
particles and antiparticles in any way. A somewhat interesting result
obtained more recently [15] has shown that the two-valuedness of photon
polarizations can be reinstated from the existence of representation spaces\
which are spanned by pairs of eigenvectors of the linear-momentum generators
of $\mathcal{P}_{+}^{\uparrow }$. It seems to have put some emphasis on the
quantum mechanical legitimacy of the twofold description of photons
mentioned anteriorly. The possibility of designing a dynamical framework
that might describe free particles and antiparticles in a unified manner has
indeed remained absent over the years from all the standard particle schemes.

In the present paper, we propose a finite-dimensional pseudo-unitary
approach to describing the dynamics of free elementary particles in a purely
relativistic quantum mechanical way. One of our postulates takes the spaces
of state vectors for any free particles or antiparticles as the
two-complex-dimensional spaces that occur in local orthogonal decompositions
of isomorphic copies of Cartan's space, the four-valued representation space
of the restricted conformal group $\mathcal{C}_{+}^{\uparrow }$ of Minkowski
space [16-18]. The implementation of this requirement relies crucially upon
the existence [16] of fundamental symmetries for Cartan's space and its
first adjoint, which hereby unites the dynamical descriptions of free
particles and antiparticles in a supposedly natural fashion. Thus, any
orthogonal direct sum of states describes locally a well-specified
particle-antiparticle pair, and carries pieces which must be normalized with
respect to the corresponding Hilbert inner products in order to fulfill a
Born-like probabilistic rule. A formal adaptation of the non-relativistic
definitions of density operators and von Neumann entropies [19] as well as
the usual prescriptions for tensor and Kronecker products, are naively
applicable to composite states of non-interacting particles and
antiparticles. All dynamical variables are taken to carry a symbolic
coordinate-free character, and thence to operate linearly on spaces of
states independently of the action of any generator of $\mathcal{C}%
_{+}^{\uparrow }$. Each of these variables amounts to an operator
restriction that has one of the two pieces borne by the orthogonal
decomposition of a copy of Cartan's space as its principal invariant
eigenspace, with the other piece being considered as the zero subspace of
the variable in question. Orbital angular momenta are assumed to be absent
from dynamical sets, whence only spin-helicity and polarization degrees of
freedom must be accounted for as far as the angular-momentum contributions
to the eventual preparations of states are concerned. Hence, electric and
all the other flavour-colour charges, spins, helicities, polarizations and
total energies are the quantities which constitute the significant complete
sets of commuting observables. The energy spectra for any massive or
massless states are at the outset incorporated into twofold patterns. In the
massive case, the values of total energies will then absorb those of linear
momenta and rest masses. Any measurements have to be performed in frames
represented univoquely by elements of $\mathcal{L}_{+}^{\uparrow }$, but
observational correlations are expressed in terms of configurations that
come from the action on spaces of state vectors of the $\mathcal{P}%
_{+}^{\uparrow }$-subgroup of an adequate realization of $SU(2,2)$.
Spacetime observers may thus keep track in a manifestly covariant way of the
behaviours of amplitudes and basis states for particles and antiparticles
that are taken away from each other along space-like or future-past
time-like directions. The local evolution of any state is controlled by a
unitary operator restriction which has to be required to depend explicitly
only upon the proper time. It follows that evolution operator equations may
be written out locally as proper-time statements. Since the spacetime
operation $PT$ does not bear an orthocronous character, it does not occur in 
$\mathcal{P}_{+}^{\uparrow }$ whence no representation of full spacetime
inversions may actually enter our descriptive framework. The dynamics of any
charged particle-antiparticle pair will rather involve the local
introduction of a conjugation operator for each charge, which is specified
together with the corresponding spin-polarization and energy spectra as a
peculiar one-to-one mapping between the pieces of the orthogonal
decomposition that defines the states for the aforesaid pair.

The work to be presented here deals with the most basic part of a programme
[16] which was initially aimed at establishing $\mathcal{P}_{+}^{\uparrow }$%
-covariant observational prescriptions for particles and antiparticles. A
basis for it comes from the belief [17] that any quantum mechanical
description of free elementary particles must be embodied into an inherently
relativistic theory which should be formulated symbolically. The operator
character of any dynamical variables thus arises essentially from the
disturbance hypothesis traditionally associated with atomic measurement
processes [20]. Presumably, the whole approach will afford a realistic
theory of free relativistic quantum mechanical twofold systems in flat
spacetime, according to which the relationships between the conformal
symmetry and null Minkowskian structures are thought of as playing no
important role. With regard to the role of $\mathcal{C}_{+}^{\uparrow }$, in
effect, the only meaning of it is related to its supply of pseudo-unitary
observational correlations within a two-level context that involves formally
[16] the maximal extension of the spacetime symmetry borne by the Wigner
schemes. One of the main outlooks we have devised from the work brings
forward the possibility of constructing methodically a background to a new
relativistic domain of quantum information theory.

The presentation has been divided into eight Sections and outlined as
follows. In Section 2, we construct the operator restrictions of interest
together with a set of formal completeness relations that will pave the way
for introducing in Section 3 the spaces of state vectors for
particle-antiparticle pairs, the apposite dynamical operators and their
local matrix representation. For the sake of organization, we shall recall
in Section 2 some of the geometric properties of Cartan's space. It will be
necessary to bring in their adjoint counterparts as well because much of our
approach unavoidably interweaves all of them. There, in Sections 2 and 3,
the representation theory developed in Ref. [17] will be taken for granted.
In Section 4, we define typical density operators and entropies. The
measurement operators which are of immediate relevance to us and the
description of their measurement processes, are exhibited in Section 5. An
appropriate description of $\mathcal{P}_{+}^{\uparrow }$ is provided in
Section 6. The observational correlations are shown in Section 7. Some
remarks on the physical contents of the work are made in Section 8.

We will adhere to the index conventions of Ref. [16]. In Sections 2 through
7, there will occur a reduction of operator representations which entails
relabelling all the components and matrix entries. The summation convention
shall be adopted unless otherwise stated explicitly. Operators will broadly
be denoted by Greek and Latin letters. A horizontal bar lying over an
indexed kernel letter will stand for the operation of complex conjugation.
The ordinary Hermitian conjugation will be indicated by a dagger
superscript. Use will sometimes be made of the natural system of units where 
$c=\hbar =k=1$, with $k$ being the Boltzmann constant. We shall also allow
for the Minkowskian signature $(+---)$. It will be convenient to adapt to
our context a bra-ket notation that interchanges the positions of the bras
and kets of any Dirac-like inner products. Bra-ket patterns carrying double
or single angular brackets will denote Hilbert or indefinite inner products,
respectively. In case an operator occurs in a bra-ket product, the vector on
which it acts will bear a single bar and the specification of its action
will be made up by attaching a double bar to the other vector. For instance,%
\begin{equation*}
<<\bullet \mid A\parallel \bullet >>.
\end{equation*}%
When there are operators acting on both sides of a bra-ket product, the
action selection will be stipulated by inserting a vertical single bar
between the desired operator blocks. As an example, we have%
\begin{equation*}
<<\bullet \mid ABC\mid DE\mid \bullet >>.
\end{equation*}%
If the blocks $ABC$ and $DE$ are one at a time taken as the identity
operator, the above product becomes%
\begin{equation*}
<<\bullet \parallel DE\mid \bullet >>,
\end{equation*}%
and%
\begin{equation*}
<<\bullet \mid ABC\parallel \bullet >>.
\end{equation*}%
If the remaining block of either case is now set as the identity operator,
we will write%
\begin{equation*}
<<\bullet \mid \bullet >>.
\end{equation*}%
This modified bra-ket notation will facilitate setting out the spectral and
measurement configurations. Further conventions will be explained in due
course.

\section{Operator restrictions and completeness relations}

Let us consider the Hilbert space $\mathcal{H}=(\mathbf{C}^{4},\mathcal{D}%
_{I})$, with $\mathcal{D}_{I}$ being the usual positive-definite inner
product on $\mathbf{C}^{4}$. By setting the canonical basis for $\mathbf{C}%
^{4}$ as $\{<e_{(\mu )}\mid \}$, and taking up linear combinations of the
form%
\begin{equation}
<\Lambda \mid =\Lambda ^{\mu }<e_{(\mu )}\mid ,\text{ }\mid \Lambda >=\mid
e_{(\nu )}>\overline{\Lambda ^{\nu }},  \tag{2.1}
\end{equation}%
we write%
\begin{equation}
\mathcal{D}_{I}{\large (}<\Phi \mid ,<\Psi \mid {\large )}\doteqdot <<\Phi
\mid \Psi >>=\hspace{0.04cm}\Phi ^{\mu }\Delta _{\mu \nu }\overline{\Psi
^{\nu }},  \tag{2.2}
\end{equation}%
where $(\Delta _{\mu \nu })$ thus denotes the identity $(4\times 4)$-matrix
whose entries are formally expressed as%
\begin{equation}
\Delta _{\mu \nu }=<<e_{(\mu )}\mid e_{(\nu )}>>.  \tag{2.3}
\end{equation}

Cartan's space shows up as the pair $\mathfrak{C}=(\mathbf{C}^{4},\mathcal{I}%
_{g})$, with $g$ being a completely invertible linear operator in $\mathbf{C}%
^{4}$ which is specified by%
\begin{equation}
g:(\Lambda ^{0},\Lambda ^{1},\Lambda ^{2},\Lambda ^{3})\mapsto (\Lambda
^{0},\Lambda ^{1},-\hspace{1pt}\Lambda ^{2},-\hspace{1pt}\Lambda ^{3}), 
\tag{2.4}
\end{equation}%
whence $g=g^{-1}$ throughout $\mathfrak{C}$. By definition, the symbol $%
\mathcal{I}_{g}$ stands for the indefinite inner product given by%
\begin{equation}
<<\Phi \mid g\parallel \Psi >>\doteqdot <\Phi \mid \Psi >_{g}=\Phi ^{\mu
}g_{\mu \nu }\overline{\Psi ^{\nu }},  \tag{2.5}
\end{equation}%
where%
\begin{equation}
g_{\mu \nu }=<e_{(\mu )}\mid e_{(\nu )}>_{g},\text{ }(g_{\mu \nu })=%
\begin{pmatrix}
I_{2} & 0_{2} \\ 
0_{2} & -I_{2}%
\end{pmatrix}%
,  \tag{2.6}
\end{equation}%
with the symbols $0_{2}$ and $I_{2}$ denoting here as elsewhere the zero and
identity $(2\times 2)$-matrices. The operator $g$ bears Hermiticity and
unitarity with respect to $\mathcal{D}_{I}$ as well as pseudo Hermiticity
and pseudo unitarity with respect to $\mathcal{I}_{g}$. We have the
inner-product correlation%
\begin{equation}
<<\Phi \mid \Psi >>=<\Phi \mid g\parallel \Psi >_{g}.  \tag{2.7}
\end{equation}%
The group $SU(2,2)$ acts transitively on $\mathfrak{C}$ as a
fifteen-real-parameter group of linear transformations that leave the
expressions (2.4)-(2.6) invariant. Any basis that satisfies (2.6) is thus
related to $\{<e_{(\mu )}\mid \}$ through an $SU(2,2)$-transformation.

One of the main geometric properties of $\mathfrak{C}$ concerns the
existence of pairs of $g$-orthogonal pseudo-Hermitian projectors in it. To
any pair $(P^{+},P^{-})$ of such projectors, there corresponds a direct-sum
splitting like [21, 23]%
\begin{equation}
\mathfrak{C}=\mathfrak{C}^{+}\oplus \mathfrak{C}^{-},  \tag{2.8}
\end{equation}%
where%
\begin{equation}
\mathfrak{C}^{\pm }\doteqdot \{<\Lambda \mid \in \mathfrak{C}:<\Lambda \mid
\Lambda >_{g}\gtrless 0\text{ \hspace{0.04cm}or }<\Lambda \mid =<0\mid \}. 
\tag{2.9}
\end{equation}%
We then have%
\begin{equation}
<\Lambda \mid =<\Lambda ^{+}\mid +<\Lambda ^{-}\mid ,\text{ }<\Lambda \mid
P^{\pm }=<\Lambda ^{\pm }\mid \in \mathfrak{C}^{\pm },  \tag{2.10}
\end{equation}%
and%
\begin{equation}
<\Lambda \mid P^{\pm }\parallel \Lambda >_{g}=<\Lambda \parallel P^{\pm
}\mid \Lambda >_{g},  \tag{2.11}
\end{equation}%
such that\footnote{%
Explicitly, $<\Lambda ^{+}\mid =(\Lambda ^{0},\Lambda ^{1},0,0)$ and $%
<\Lambda ^{-}\mid =(0,0,\Lambda ^{2},\Lambda ^{3})$.}%
\begin{equation}
<e_{(\mu )}\mid P^{\pm }=<e_{(\mu )}^{\pm }\mid ,\text{ }<\Lambda ^{\pm
}\mid =\Lambda ^{\mu }<e_{(\mu )}^{\pm }\mid .  \tag{2.12}
\end{equation}%
Hence, we can reexpress $\mathcal{I}_{g}$ as either of the prescriptions%
\begin{equation}
<\Phi \mid \Psi >_{g}=<<\Phi ^{+}\mid \Psi ^{+}>>-<<\Phi ^{-}\mid \Psi
^{-}>>,  \tag{2.13a}
\end{equation}%
and%
\begin{equation}
g_{\mu \nu }=<<e_{(\mu )}^{+}\mid e_{(\nu )}^{+}>>-<<e_{(\mu )}^{-}\mid
e_{(\nu )}^{-}>>.  \tag{2.13b}
\end{equation}

For an invertible linear operator $A$ in $\mathfrak{C}$, we have the
expansions%
\begin{equation}
<\Lambda \mid AP^{+}=<\Lambda ^{+}\mid A^{++}+<\Lambda ^{-}\mid A^{-+}, 
\tag{2.14a}
\end{equation}%
and%
\begin{equation}
<\Lambda \mid AP^{-}=<\Lambda ^{+}\mid A^{+-}+<\Lambda ^{-}\mid A^{--}, 
\tag{2.14b}
\end{equation}%
together with the scheme%
\begin{equation}
\begin{array}{c}
A^{++}:\mathfrak{C}^{+}\rightarrow R(A)\cap \mathfrak{C}^{+},\text{ }A^{+-}:%
\mathfrak{C}^{+}\rightarrow R(A)\cap \mathfrak{C}^{-}, \\ 
A^{-+}:\mathfrak{C}^{-}\rightarrow R(A)\cap \mathfrak{C}^{+},\text{ }A^{--}:%
\mathfrak{C}^{-}\rightarrow R(A)\cap \mathfrak{C}^{-}.%
\end{array}
\tag{2.15}
\end{equation}%
A formal decomposition of the matrix elements%
\begin{equation}
A_{\mu \nu }=<e_{(\mu )}\mid A\parallel e_{(\nu )}>_{g},  \tag{2.16}
\end{equation}%
thus emerges out of implementing (2.15). This procedure leads us, in effect,
to the following block representation for the operator $A$:%
\begin{equation}
(A_{\mu \nu })=%
\begin{pmatrix}
<e_{(\mu )}^{+}\mid A^{++}\parallel e_{(\nu )}^{+}>_{g^{+}} & <e_{(\mu
)}^{+}\mid A^{+-}\parallel e_{(\nu )}^{-}>_{g^{-}} \\ 
<e_{(\mu )}^{-}\mid A^{-+}\parallel e_{(\nu )}^{+}>_{g^{+}} & <e_{(\mu
)}^{-}\mid A^{--}\parallel e_{(\nu )}^{-}>_{g^{-}}%
\end{pmatrix}%
.  \tag{2.17}
\end{equation}

We recall [23] that a subspace $\mathfrak{L}$ of $\mathfrak{C}$ is said to
be an invariant subspace of $A$ if the range $A(\mathfrak{C}\cap \mathfrak{L}%
)$ is contained in $\mathfrak{L}$. A restriction Res$A_{\mathfrak{L}}$ of $A$
on such an $\mathfrak{L}$ is an invertible linear operator whose domain and
range are prescribed as%
\begin{equation}
D(\text{Res}A_{\mathfrak{L}})=\mathfrak{C}\cap \mathfrak{L},\text{ }R(\text{%
Res}A_{\mathfrak{L}})\subset \mathfrak{L}\subset R(A).  \tag{2.18}
\end{equation}%
If $\mathfrak{M}$ is another invariant subspace of $A$ such that $\mathfrak{L%
}\cap \mathfrak{M}=<0\mid $, and the inner product on $\mathfrak{C}$ does
not degenerate on both $\mathfrak{L}$ and $\mathfrak{M}$, then we may write
the splittings%
\begin{equation}
\mathfrak{C}=\mathfrak{L}\oplus \mathfrak{M}\Leftrightarrow D(A)=D(\text{Res}%
A_{\mathfrak{L}})\oplus D(\text{Res}A_{\mathfrak{M}}),  \tag{2.19a}
\end{equation}%
and%
\begin{equation}
A=\text{Res}A_{\mathfrak{L}}\oplus \text{Res}A_{\mathfrak{M}},  \tag{2.19b}
\end{equation}%
along with the defining zero subspaces%
\begin{equation}
\mathfrak{N}(\text{Res}A_{\mathfrak{L}})\doteqdot \mathfrak{M},\text{ }%
\mathfrak{N}(\text{Res}A_{\mathfrak{M}})\doteqdot \mathfrak{L}.  \tag{2.20}
\end{equation}%
When $A$ is restricted to $\mathfrak{C}^{\pm }$, we obtain the operator array%
\footnote{%
In Section 3, the ranges of restrictions will be identified with the
respective invariant subspaces.}%
\begin{equation}
A=\left( 
\begin{array}{ll}
\text{Res}A_{\mathfrak{C}^{+}} & 0 \\ 
0 & \text{Res}A_{\mathfrak{C}^{-}}%
\end{array}%
\right) ,  \tag{2.21}
\end{equation}%
together with the restricted expansion%
\begin{equation}
<\Lambda \mid A=<\Lambda ^{+}\mid \text{Res}A_{\mathfrak{C}^{+}}+<\Lambda
^{-}\mid \text{Res}A_{\mathfrak{C}^{-}},  \tag{2.22}
\end{equation}%
and the definitions%
\begin{equation}
D(\text{Res}A_{\mathfrak{C}^{\pm }})=\mathfrak{C}^{\pm }\supseteq R(\text{Res%
}A_{\mathfrak{C}^{\pm }}),\text{ }\mathfrak{N}(\text{Res}A_{\mathfrak{C}%
^{\pm }})=\mathfrak{C}^{\mp }.  \tag{2.23}
\end{equation}%
We can therefore account for the property%
\begin{equation}
\text{Res}(ABC)_{\mathfrak{C}^{\pm }}=\text{Res}A_{\mathfrak{C}^{\pm }}\text{%
Res}B_{\mathfrak{C}^{\pm }}\text{Res}C_{\mathfrak{C}^{\pm }}.  \tag{2.24}
\end{equation}

The restrictions of the identity operator $I$ on $\mathfrak{C}$ lead to the
representative entries%
\begin{equation}
g_{\mu \nu }^{\pm }=<e_{(\mu )}^{\pm }\mid \text{Res}I_{\mathfrak{C}^{\pm
}}\parallel e_{(\nu )}^{\pm }>_{g^{\pm }}=\pm <<e_{(\mu )}^{\pm }\mid
e_{(\nu )}^{\pm }>>,  \tag{2.25}
\end{equation}%
which, in turn, yield the reduced $(2\times 2)$-block matrices%
\begin{equation}
(g_{\mu \nu }^{+})=\left( 
\begin{array}{ll}
1 & 0 \\ 
0 & 1%
\end{array}%
\right) ,\text{ }(g_{\mu \nu }^{-})=\left( 
\begin{array}{ll}
-1 & 0 \\ 
0 & -1%
\end{array}%
\right) ,  \tag{2.26}
\end{equation}%
with $g^{\pm }\doteqdot $ Res$g_{\mathfrak{C}^{\pm }}$. It is worth pointing
out that any $g$-restriction is allowed to be defined only on $\mathfrak{C}%
^{\pm }$. Hence, introducing the reduced component setting%
\begin{equation}
\Lambda ^{0}=\Lambda _{+}^{0},\text{ }\Lambda ^{1}=\Lambda _{+}^{1},\text{ }%
\Lambda ^{2}=\Lambda _{-}^{0},\text{ }\Lambda ^{3}=\Lambda _{-}^{1}, 
\tag{2.27a}
\end{equation}%
and rewriting the second of (2.12) as%
\begin{equation}
<\Lambda ^{\pm }\mid =\Lambda _{\pm }^{\mu }<e_{(\mu )}^{\pm }\mid , 
\tag{2.27b}
\end{equation}%
we recast the\ inner products on $\mathfrak{C}^{\pm }$ into the definite
configuration%
\begin{equation}
\mathcal{I}_{g^{\pm }}{\large (}<\Phi ^{\pm }\mid ,<\Psi ^{\pm }\mid {\large %
)}\doteqdot <\Phi ^{\pm }\mid \Psi ^{\pm }>_{g^{\pm }}=\Phi _{\pm }^{\mu
}g_{\mu \nu }^{\pm }\overline{\Psi _{\pm }^{\nu }},  \tag{2.28}
\end{equation}%
which involves the restricted pseudo-Hermiticity relation%
\begin{equation}
<\Phi ^{\pm }\mid g^{\pm }\parallel \Psi ^{\pm }>_{g^{\pm }}=<\Phi ^{\pm
}\parallel g^{\pm }\mid \Psi ^{\pm }>_{g^{\pm }}.  \tag{2.29}
\end{equation}%
It follows that%
\begin{equation}
<e_{(\mu )}\mid P^{\pm }\parallel e_{(\nu )}>_{g}=g_{\mu \nu }^{\pm }, 
\tag{2.30}
\end{equation}%
whereas the representation of Res$I_{\mathfrak{C}^{\pm }}$ with respect to $%
\mathcal{D}_{I}$ is constituted by the entries%
\begin{equation}
<<e_{(\mu )}^{\pm }\mid \text{Res}I_{\mathfrak{C}^{\pm }}\parallel e_{(\nu
)}^{\pm }>>=\Delta _{\mu \nu }^{\pm },\text{ }  \tag{2.31}
\end{equation}%
which give the reduced matrices%
\begin{equation}
(\Delta _{\mu \nu }^{+})=\left( 
\begin{array}{ll}
1 & 0 \\ 
0 & 1%
\end{array}%
\right) =(\Delta _{\mu \nu }^{-}).  \tag{2.32}
\end{equation}

The first adjoint $\mathfrak{C}^{\ast }$ of $\mathfrak{C}$ is defined in
such a way that each element of $\mathfrak{C}$ enters a one-to-one mapping
which produces the basis relationships [16]%
\begin{equation}
<e_{(\mu )}^{\pm }\mid \mapsto g_{\mu \lambda }^{\pm }<e_{\pm }^{\ast
(\lambda )}\mid ,\text{ }<e_{\pm }^{\ast (\mu )}\mid \mapsto g_{\pm }^{\ast
\mu \lambda }<e_{(\lambda )}^{\pm }\mid ,  \tag{2.33}
\end{equation}%
whence it is legitimate to write down the splitting%
\begin{equation}
\mathfrak{C}^{\ast }=\mathfrak{C}_{+}^{\ast }\oplus \mathfrak{C}_{-}^{\ast }.
\tag{2.34}
\end{equation}%
Of course, the operator rule for $g^{\ast }$ is formally the same as the one
for $g$. Therefore, for some element of $\mathfrak{C}^{\ast }$, we can take
account of the adjoint-component identifications%
\begin{equation}
\Lambda _{0}=\Lambda _{0}^{+},\text{ }\Lambda _{1}=\Lambda _{1}^{+},\text{ }%
\Lambda _{2}=\Lambda _{0}^{-},\text{ }\Lambda _{3}=\Lambda _{1}^{-}, 
\tag{2.35}
\end{equation}%
to spell out the reduced configurations%
\begin{equation}
<\Lambda _{\pm }^{\ast }\mid =\Lambda _{\mu }^{\pm }<e_{\pm }^{\ast (\mu
)}\mid ,\text{ }\Lambda _{\mu }^{\pm }=\Lambda _{\pm }^{\lambda }g_{\lambda
\mu }^{\pm },\text{ }\Lambda _{\pm }^{\mu }=\Lambda _{\lambda }^{\pm }g_{\pm
}^{\ast \lambda \mu },  \tag{2.36}
\end{equation}%
which yield the expressions\footnote{%
It should be clear that $\mathfrak{C}^{\pm }\simeq \mathbf{C}^{2}\simeq 
\mathfrak{C}_{\pm }^{\ast }$. Accordingly, the entry labels borne by any
reduced or restricted structures must take the values $0$ and $1$.}%
\begin{equation}
<\Lambda ^{\pm }\mid e_{(\mu )}^{\pm }>_{g^{\pm }}=\Lambda _{\mu }^{\pm },%
\text{ }<\Lambda _{\pm }^{\ast }\mid e_{\pm }^{\ast (\mu )}>_{g_{\pm }^{\ast
}}=\Lambda _{\pm }^{\mu }.  \tag{2.37}
\end{equation}

The relationships (2.33) do not depend upon the choice of $<\Lambda \mid $,
but they carry forward the canonical character of $<e_{(\mu )}\mid $ to $%
<e^{\ast (\mu )}\mid $. Hence, the entries%
\begin{equation}
\Delta _{\pm }^{\ast \mu \nu }=<<e_{\pm }^{\ast (\mu )}\mid e_{\pm }^{\ast
(\nu )}>>,\text{ }g_{\pm }^{\ast \mu \nu }=<e_{\pm }^{\ast (\mu )}\mid
e_{\pm }^{\ast (\nu )}>_{g_{\pm }^{\ast }},  \tag{2.38}
\end{equation}%
coincide with those of $(\Delta _{\mu \nu }^{\pm })$ and $(g_{\mu \nu }^{\pm
})$, respectively. Furthermore, such entries satisfy the correlations%
\begin{equation}
g_{\mu \nu }^{\pm }=\Delta _{\mu \lambda }^{\pm }g_{\pm }^{\ast \lambda
\sigma }\Delta _{\sigma \nu }^{\pm },\text{ }g_{\pm }^{\ast \mu \nu }=\Delta
_{\pm }^{\ast \mu \lambda }g_{\lambda \sigma }^{\pm }\Delta _{\pm }^{\ast
\sigma \nu },  \tag{2.39}
\end{equation}%
along with the ones that are obtained from (2.39) by interchanging the
kernel letters $\Delta $ and $g$. It is shown in Ref. [16] that the
realizations of $SU(2,2)$ in $\mathfrak{C}$ and $\mathfrak{C}^{\ast }$ are
just the same. Any of these realizations is constituted by the set of
complex $(4\times 4)$-matrices which arise from the representation of
pseudo-unitary operators in $\mathfrak{C}$ and $\mathfrak{C}^{\ast }$ (for
further details, see Ref. [17]; see also Section 6).

For the restriction carried by (2.25), we have the decomposition%
\begin{equation}
\text{Res}I_{\mathfrak{C}^{\pm }}=\mid e_{(\mu )}^{\pm }>g_{\pm }^{\ast \mu
\nu }<e_{(\nu )}^{\pm }\mid ,  \tag{2.40}
\end{equation}%
which shows us that $<e_{(\mu )}^{\pm }\mid $ possesses a completeness
property. So by adapting to $A$ the notation of (2.26), and likewise taking
into account the adjoint of (2.16), we write the restrictions%
\begin{equation}
A^{\pm }=\mid e_{(\mu )}^{\pm }>A_{\pm }^{\ast \mu \nu }<e_{(\nu )}^{\pm
}\mid ,\text{ }A_{\pm }^{\ast }=\mid e_{\pm }^{\ast (\mu )}>A_{\mu \nu
}^{\pm }<e_{\pm }^{\ast (\nu )}\mid ,  \tag{2.41a}
\end{equation}%
along with the entry relations [17]%
\begin{equation}
A_{\mu \nu }^{\pm }=g_{\mu \lambda }^{\pm }A_{\pm }^{\ast \lambda \sigma
}g_{\sigma \nu }^{\pm },\text{ }A_{\pm }^{\ast \mu \nu }=g_{\pm }^{\ast \mu
\lambda }A_{\lambda \sigma }^{\pm }g_{\pm }^{\ast \sigma \nu }.  \tag{2.41b}
\end{equation}%
For the trace of $A^{\pm }$, we have the expression%
\begin{equation}
\text{Tr }A^{\pm }=<e_{(\mu )}^{\pm }\mid A^{\pm }\parallel e_{(\nu )}^{\pm
}>_{g^{\pm }}g_{\pm }^{\ast \nu \mu }.  \tag{2.42}
\end{equation}

It is useful to define an operator associated to (2.12) as the projector%
\begin{equation}
\mid \Lambda ^{\pm }><\Lambda ^{\pm }\mid =\mid e_{(\mu )}^{\pm }>\overline{%
\Lambda _{\pm }^{\mu }}\Lambda _{\pm }^{\nu }<e_{(\nu )}^{\pm }\mid . 
\tag{2.43}
\end{equation}%
Structures of this kind shall be used in Section 4 for defining density
restrictions. When coupling (2.43) to other operators, it will generally be
convenient to account for the double-bar convention explained in Section 1.
From (2.42), we thus get the trace%
\begin{equation}
\text{Tr}\mid \Lambda ^{\pm }><\Lambda ^{\pm }\mid =<\Lambda ^{\pm }\mid
\Lambda ^{\pm }>_{g^{\pm }},  \tag{2.44}
\end{equation}%
together with the projections%
\begin{equation}
\mid \Lambda ^{\pm }><\Lambda ^{\pm }\mid e_{(\mu )}^{\pm }>_{g^{\pm }}=\mid
\Lambda ^{\pm }>\Lambda _{\mu }^{\pm },  \tag{2.45a}
\end{equation}%
and%
\begin{equation}
<e_{(\mu )}^{\pm }\mid \Lambda ^{\pm }>_{g}<\Lambda ^{\pm }\mid =\overline{%
\Lambda _{\mu }^{\pm }}<\Lambda ^{\pm }\mid .  \tag{2.45b}
\end{equation}

\section{States, dynamical variables and local spectra}

One of the key attitudes towards settling down our dynamical approach,
starts taking a copy of $\mathfrak{C}^{\pm }$ or else of $\mathfrak{C}_{\pm
}^{\ast }$ as the space of state vectors for any free elementary particle or
antiparticle, regardless of whether the system under consideration is a
fermion or a boson. Of course, this procedure goes hand-in-hand with the
Pauli-Weisskopf theorem [27] which states that the occurrence in nature of
particles and antiparticles does not depend upon any spin values. Thus, the
dynamical states of every particle belong to a copy of $\mathfrak{C}^{+}$ or 
$\mathfrak{C}_{+}^{\ast }$ whereas the states of every antiparticle belong
to a copy of $\mathfrak{C}^{-}$ or $\mathfrak{C}_{-}^{\ast }$. Evidently,
this prescription presupposes that the local observer for an element of $%
\mathcal{L}_{+}^{\uparrow }$ is chosen for which spectral preparations and
measurement processes should be performed in either case.

The observables that will take part of the descriptions to be set up
hereabout are those referred to in Section 1 to which Naimark's theorems
[24] concerning the existence in $\mathfrak{C}$ and $\mathfrak{C}^{\ast }$
of common eigenvectors for sets of commuting pseudo-Hermitian operators do
surely apply. Every particle or antiparticle is physically identified
through its complete set of observables. Any dynamical variable appears as a
pseudo-Hermitian operator restriction that takes up adequately a copy of $%
\mathfrak{C}^{\pm }$ or $\mathfrak{C}_{\pm }^{\ast }$ as its principal
invariant eigenspace, in addition to possessing a range which effectively
equals its domain. Moreover, it holds a real bounded two-valued spectrum
whose reality is formally ensured in any case [21, 23] by the definiteness
of the inner products on $\mathfrak{C}^{\pm }$ and $\mathfrak{C}_{\pm
}^{\ast }$. All charge operators possess degenerate spectra, in accordance
with the standard particle theories [8]. To see what a typical pattern of
such charge spectra looks like, it will suffice to work out the
representation for some charge. Spin and polarization spectra shall absorb
the same quantum-number prescriptions as the discrete ones that are borne by
the particle classification schemes brought up in Section 1. We should
emphasize, however, that the only fermions which may enter the dynamical
approach are identified with leptons and quarks that move freely for a while
after the occurrence of elementary processes. The same feasibility applies
to bosons too, but the dynamics of gravitons can not be incorporated into
here. Therefore, only the values $\pm 1/2$ and $\pm 1$ will partake of the
spin-polarization spectra. We stress, in particular, that energy operators
are not required to bear a coordinate character related to time translations
or any generators of $\mathcal{P}_{+}^{\uparrow }$. Instead, each of them
must just be taken as a symbolic entity that supplies a discrete spectrum
for some free particle or antiparticle.

The orthogonal decompositions that yield spaces of states suggest following
up a view which conforms to the requirement [28] that free particles and
antiparticles not only may travel in spacetime along future and past
directions while carrying positive and negative energies, but also must bear
opposite charges. States and dynamical variables for pairs of free
particle-antiparticle companions are provided by configurations like the
ones of (2.10) and (2.21). Upon being read from left to right, the entries
of each such pair will thus refer exclusively to a particle and its
antiparticle counterpart. Inasmuch as the Born rule still holds for every
individual state, the direct-sum state for any pair has to be normalized in
a characteristic manner. The charge conjugations for some charged pair are
defined as invertible linear operators that map the pertinent copies of $%
\mathfrak{C}^{+}$ and $\mathfrak{C}_{+}^{\ast }$ into the corresponding ones
of $\mathfrak{C}^{-}$ and $\mathfrak{C}_{-}^{\ast }$, respectively. A
notable property of such mappings, which differs them conceptually from the
electric-charge one borne by the ordinary relativistic context [8, 28], is
that the independence explained in Section 1 between their definition and
the actions of parity and time-reversal operators in Minkowski space,
produces spin-energy associations between particles and antiparticles that
are attainable in a fixed Lorentz frame. State vectors will not therefore be
sensitive to any improper or non-orthocronous operations in spacetime,
whence all negative energies shall be regarded as nothing else but formal
spectral constituents.

There exist unitary operators that allow changing locally the description of
degrees of freedom. These unitary methods will give rise to structures which
afford a symbolic definition of helicity operators. We will likewise see how
the use of suitably selected local bases may produce the possibility of
interchanging spin-polarization characters. The products which lead to any
up-down, vertical-horizontal and left-right spectra shall get rid of the
procedures that involve taking spin and polarization components along
locally specified $OXYZ$-axes and directions of motion.\footnote{%
Any helicity descriptions that do not bear two-foldness involve the
implementation of spurious procedures. Here, such situations will be
entirely disregarded.} As outlined before, we shall now carry out the
construction of observables and spectra. Many of the formulae exhibited in
Section 2 will be used so many times herein that we shall no longer refer to
them explicitly.

Henceforward the spaces $\mathfrak{C}$ and $\mathfrak{C}^{\ast }$ will
themselves serve as dynamical prototypes for a spinning or polarized charged
pair $(p^{+},p^{-})$. We will first write the unstarred structures for $%
(p^{+},p^{-})$ without leaving out the main adjoint counterparts. At this
point, we shall complete our procedures in the $\mathcal{L}_{+}^{\uparrow }$%
-frame that utilizes $\{<e_{(\mu )}^{\pm }\mid \}$ as the computational
basis for the pair allowed for. The local form of the patterns that should
take place when particles or antiparticles are considered alone, will become
automatically available thereafter. Any copy of the unstarred canonical
basis must constitute in the same spacetime frame a complete set of states
for some pair. It should thus be reset as a reduced device that spans some
copy of $\mathfrak{C}^{\pm }$, in accordance with the prescriptions of
Section 2. For this reason, we will also use the symbols $p^{+}$ and $p^{-}$
for labelling the adjoint computational bases for $(p^{+},p^{-})$. For
instance,\footnote{%
There will be no need in what follows to reset (3.1) as row vectors.}%
\begin{equation}
<e_{(0)}^{(p\pm )}\mid =%
\begin{pmatrix}
1 \\ 
0%
\end{pmatrix}%
,\text{ }<e_{(1)}^{(p\pm )}\mid =%
\begin{pmatrix}
0 \\ 
1%
\end{pmatrix}%
.  \tag{3.1}
\end{equation}

Any $\mathfrak{C}$-state for the pair $(p^{+},p^{-})$ possesses the form%
\begin{equation}
<\Phi ^{(p^{+}p^{-})}\mid =<\Phi ^{(p+)}\mid +<\Phi ^{(p-)}\mid ,\text{ }%
<\Phi ^{(p\pm )}\mid \in \mathfrak{C}^{\pm },  \tag{3.2}
\end{equation}%
with its pieces being normalized as%
\begin{equation}
<<\Phi ^{(p\pm )}\mid \Phi ^{(p\pm )}>>=<\Phi ^{(p\pm )}\mid g^{\pm
}\parallel \Phi ^{(p\pm )}>_{g^{\pm }}=1,  \tag{3.3}
\end{equation}%
such that%
\begin{equation}
<<\Phi ^{(p^{+}p^{-})}\mid \Phi ^{(p^{+}p^{-})}>>=2.  \tag{3.4}
\end{equation}%
Invoking (2.13) yields the characterization%
\begin{equation}
<\Phi ^{(p^{+}p^{-})}\mid \Phi ^{(p^{+}p^{-})}>_{g}=0,\text{ }<\Phi ^{(p\pm
)}\mid \Phi ^{(p\pm )}>_{g^{\pm }}=\pm 1,  \tag{3.5}
\end{equation}%
which means that every unstarred state for $(p^{+},p^{-})$ has to be taken
as a null vector with respect to $\mathcal{I}_{g}$. More explicitly, for
(3.3), we have the expression%
\begin{equation}
<<\Phi ^{(p\pm )}\mid \Phi ^{(p\pm )}>>=\hspace{0.04cm}\Phi _{(p\pm )}^{\mu
}\Delta _{\mu \nu }^{\pm }\overline{\Phi _{(p\pm )}^{\nu }}.  \tag{3.6}
\end{equation}

States of the form of (3.2) bear purity in the ordinary sense, and are
prepared so as to yield a completion of test and measurement processes for
particles and antiparticles. Such states describe pairs like $(p^{+},p^{-})$%
, and thereby do not amount to any composite states (see Section 4). The
relevant adjoint relationships between dynamical amplitudes and Born
probabilities are thus set as configurations of the type%
\begin{equation}
w_{(\mu )}^{(p\pm )}=\mid \Phi _{(p\pm )}^{\lambda }\Delta _{\lambda \mu
}^{\pm }\mid ^{2},\text{ }w_{(p\pm )}^{(\mu )}=\mid \Phi _{\lambda }^{(p\pm
)}\Delta _{\pm }^{\ast \lambda \mu }\mid ^{2},  \tag{3.7}
\end{equation}%
with%
\begin{equation}
<\Phi _{(p\pm )}^{\ast }\mid =\Phi _{\mu }^{(p\pm )}<e_{(p\pm )}^{\ast (\mu
)}\mid .  \tag{3.8}
\end{equation}%
Obviously, the values (3.7) remain unaltered when the kernel letter $\Delta $
is replaced with $g$. Therefore,%
\begin{equation}
w_{(\mu )}^{(p\pm )}=w_{(p\pm )}^{(\mu )},  \tag{3.9}
\end{equation}%
with%
\begin{equation}
w_{(0)}^{(p+)}=\mid \Phi _{(p+)}^{0}\mid ^{2},\text{ }w_{(1)}^{(p+)}=\mid
\Phi _{(p+)}^{1}\mid ^{2},  \tag{3.10a}
\end{equation}%
and%
\begin{equation}
w_{(0)}^{(p-)}=\mid \Phi _{(p-)}^{0}\mid ^{2},\text{ }w_{(1)}^{(p-)}=\mid
\Phi _{(p-)}^{1}\mid ^{2}.  \tag{3.10b}
\end{equation}%
Hence, (3.6) can be rewritten as the normalized expansion%
\begin{equation}
<<\Phi ^{(p\pm )}\mid \Phi ^{(p\pm )}>>=w_{(0)}^{(p\pm )}+w_{(1)}^{(p\pm )}.
\tag{3.11}
\end{equation}

The evolution of any states for $(p^{+},p^{-})$ is governed\footnote{%
Unitary operators that do not bear pseudo unitarity will from now onwards be
denoted by either upright Latin or Gothic letters.} by the restrictions on $%
\mathfrak{C}^{\pm }$ and $\mathfrak{C}_{\pm }^{\ast }$ of unitary operators $%
\{\mathfrak{U},\mathfrak{U}^{\ast }\}$ whose matrix representation lies
outside the special intersection%
\begin{equation}
SU(2,2)\cap U(4).  \tag{3.12}
\end{equation}%
In the unstarred case, such an evolution is brought about locally by
statements of the form%
\begin{equation}
<\Phi ^{(p\pm )}(\tau )\mid =<\Phi ^{(p\pm )}(\tau _{0})\mid \mathfrak{U}%
^{\pm }(\tau ,\tau _{0}),  \tag{3.13}
\end{equation}%
where $\tau $ stands for the proper time, and%
\begin{equation}
\mathfrak{U}(\tau ,\tau _{0})=\left( 
\begin{array}{ll}
\mathfrak{U}^{+}(\tau ,\tau _{0}) & 0 \\ 
0 & \mathfrak{U}^{-}(\tau ,\tau _{0})%
\end{array}%
\right) .  \tag{3.14}
\end{equation}%
The basis states carried by (3.13) should be held fixed as the action of $%
\mathfrak{U}^{\pm }(\tau ,\tau _{0})$ is implemented, whence the amplitudes $%
\Phi _{(p\pm )}^{\mu }$ must undergo the evolution law%
\begin{equation}
\Phi _{(p\pm )}^{\mu }(\tau )=\Phi _{(p\pm )}^{\lambda }(\tau _{0})\mathfrak{%
U}_{\lambda }^{\pm }{}^{\mu }(\tau ,\tau _{0}).  \tag{3.15}
\end{equation}

Every significant observable for some pair is a completely invertible linear
operator $A$ that consists of pseudo-Hermitian restrictions with respect to
the corresponding copy of $\mathcal{I}_{g}$, namely,%
\begin{equation}
A=\left( 
\begin{array}{ll}
A^{+} & 0 \\ 
0 & A^{-}%
\end{array}%
\right) =\left( 
\begin{array}{ll}
A^{+\bigstar } & 0 \\ 
0 & A^{-\bigstar }%
\end{array}%
\right) =A^{\bigstar }.  \tag{3.16}
\end{equation}%
The domain-range definitions for $A^{\pm }$ satisfy%
\begin{equation}
D(A^{\pm })=R(A^{\pm }),  \tag{3.17}
\end{equation}%
and the prescriptions (2.23) still specify the respective null spaces. In
Ref. [17], it was shown for the first time that any array of the form of
(3.16) obeys the relation $A^{\bigstar }=A^{\dag }$, whence all observables
must likewise bear Hermiticity. For the restrictions $A^{(p\pm )}$ of the
observable $A^{(p^{+}p^{-})}$ for $(p^{+},p^{-})$, we have the spectral
entries%
\begin{equation}
A_{\mu \nu }^{(p\pm )}\doteqdot A_{\mu }^{(p\pm )\lambda }g_{\lambda \nu
}^{\pm }=<e_{(\mu )}^{(p\pm )}\mid A^{(p\pm )}\parallel e_{(\nu )}^{(p\pm
)}>_{g^{\pm }},  \tag{3.18a}
\end{equation}%
whose adjoint version is%
\begin{equation}
A_{(p\pm )}^{\ast \mu \nu }\doteqdot A_{(p\pm )\lambda }^{\ast \mu }g_{\pm
}^{\ast \lambda \nu }=<e_{(p\pm )}^{\ast (\mu )}\mid A_{(p\pm )}^{\ast
}\parallel e_{(p\pm )}^{\ast (\nu )}>_{g_{\pm }^{\ast }}.  \tag{3.18b}
\end{equation}%
Any entries like $A_{\mu }^{(p\pm )\nu }$ and $A_{(p\pm )\nu }^{\ast \mu }$
carry an intrinsic character\ such that, for some given local bases, their
values do not depend upon which inner products are occasionally implemented.

We can invoke the definiteness of the inner products (2.28) to translate
formally the spectra of $A^{(p\pm )}$ and $A_{(p\pm )}^{\ast }$ into the
real reduced matrices%
\begin{equation}
(A_{\mu \nu }^{(p\pm )})=\left( 
\begin{array}{ll}
A_{00}^{(p\pm )} & 0 \\ 
0 & A_{11}^{(p\pm )}%
\end{array}%
\right) ,\text{ }(A_{(p\pm )}^{\ast \mu \nu })=\left( 
\begin{array}{ll}
A_{(p\pm )}^{\ast 00} & 0 \\ 
0 & A_{(p\pm )}^{\ast 11}%
\end{array}%
\right) ,  \tag{3.19}
\end{equation}%
which are equal to one another because of (2.41b). Hence, utilizing (2.42)
together with its adjoint, gives the traces%
\begin{equation}
\text{Tr }A^{(p\pm )}=\pm A_{00}^{(p\pm )}\pm A_{11}^{(p\pm )},\text{ Tr }%
A_{(p\pm )}^{\ast }=\pm A_{(p\pm )}^{\ast 00}\pm A_{(p\pm )}^{\ast 11}. 
\tag{3.20}
\end{equation}%
The expectation value of $A^{(p\pm )}$ in the $\Phi $-state is then
expressed by%
\begin{equation}
<A^{(p\pm )}>_{\Phi ^{\pm }}=<\Phi ^{(p\pm )}\mid A^{(p\pm )}\parallel \Phi
^{(p\pm )}>_{g^{\pm }}=\Phi _{(p\pm )}^{\mu }A_{\mu \nu }^{(p\pm )}\overline{%
\Phi _{(p\pm )}^{\nu }},  \tag{3.21}
\end{equation}%
which evidently satisfies (2.41a). For any restrictions $R^{(p\pm )}$ that
possess a dynamical significance, it follows that we can write locally
proper-time evolution statements like%
\begin{equation}
R^{(p\pm )}(\tau )=\mathfrak{U}^{\pm }{}(\tau ,\tau _{0})R^{(p\pm )}(\tau
_{0})\mathfrak{U}^{\pm \dag }{}(\tau ,\tau _{0}),  \tag{3.22}
\end{equation}%
with $\mathfrak{U}^{\pm }{}(\tau ,\tau _{0})$ bearing the same meaning as
before.

In fact, the observables involved in the dynamics of any pair commute with
each other. Therefore, for two such observables, the definitions (2.23)
produce the local commutator statement%
\begin{equation}
\lbrack A,B]=0\Leftrightarrow \lbrack A^{\pm },B^{\pm }]=0.  \tag{3.23}
\end{equation}%
For $(p^{+},p^{-})$, we thus have the eigenvalue equation%
\begin{equation}
<e_{(\mu )}^{(p\pm )}\mid A^{(p\pm )}=a_{\mu }^{(p\pm )}<e_{(\mu )}^{(p\pm
)}\mid \text{(no summation over here)},  \tag{3.24}
\end{equation}%
which, consequently, holds formally for any of the other observables for $%
(p^{+},p^{-})$. Equation (3.24) yields the product%
\begin{equation}
<e_{(\mu )}^{(p\pm )}\mid A^{(p\pm )}\parallel e_{(\nu )}^{(p\pm )}>_{g^{\pm
}}=a_{\mu }^{(p\pm )}g_{\mu \nu }^{\pm }\text{ (no summation over here)}, 
\tag{3.25}
\end{equation}%
which, in view of (2.41a), gives rise to the spectral decompositions%
\begin{equation}
A^{(p\pm )}=\pm \mid e_{(0)}^{(p\pm )}>a_{(p\pm )}^{0}<e_{(0)}^{(p\pm )}\mid
\pm \mid e_{(1)}^{(p\pm )}>a_{(p\pm )}^{1}<e_{(1)}^{(p\pm )}\mid . 
\tag{3.26}
\end{equation}%
The matrices of (3.19) can then be reexpressed as the adjoint configurations 
\begin{equation}
(A_{\mu \nu }^{(p\pm )})=\left( 
\begin{array}{ll}
\pm a_{0}^{(p\pm )} & 0 \\ 
0 & \pm a_{1}^{(p\pm )}%
\end{array}%
\right) ,\text{ }(A_{(p\pm )}^{\ast \mu \nu })=\left( 
\begin{array}{ll}
\pm a_{(p\pm )}^{0} & 0 \\ 
0 & \pm a_{(p\pm )}^{1}%
\end{array}%
\right) .  \tag{3.27}
\end{equation}%
Whence, the value (3.21) may be given by the reduced formula%
\begin{equation}
<A^{(p\pm )}>_{\Phi ^{\pm }}=\pm a_{\mu }^{(p\pm )}\mid \Phi _{(p\pm )}^{\mu
}\mid ^{2},  \tag{3.28}
\end{equation}%
whereas (3.20) becomes%
\begin{equation}
\text{Tr }A^{(p\pm )}=a_{0}^{(p\pm )}+a_{1}^{(p\pm )},\text{ Tr }A_{(p\pm
)}^{\ast }=a_{(p\pm )}^{0}+a_{(p\pm )}^{1}.  \tag{3.29}
\end{equation}%
In many cases where $f(A^{(p\pm )})$ is employed in place of $A^{(p\pm )}$,
we may substitute $f(a_{\mu }^{(p\pm )})$ for $a_{\mu }^{(p\pm )}$. This
will be used in Section 4 for computing some entropic values.

The restrictions for a typical charge operator $Q^{(p^{+}p^{-})}$ for $%
(p^{+},p^{-})$ lead to the eigenvalue equations%
\begin{equation}
<e_{(\mu )}^{(p\pm )}\mid Q^{(p\pm )}=\pm q<e_{(\mu )}^{(p\pm )}\mid , 
\tag{3.30}
\end{equation}%
along with the degenerate reduced spectra%
\begin{equation}
(Q_{\mu \nu }^{(p\pm )})=\left( 
\begin{array}{ll}
q & 0 \\ 
0 & q%
\end{array}%
\right) =(Q_{(p\pm )}^{\ast \mu \nu }),  \tag{3.31}
\end{equation}%
where $q$ and $-q$ stand for the corresponding charges of the particle $%
p^{+} $and its companion $p^{-}$, respectively. Any charge conjugations are
defined by dimensionless linear operators of the form%
\begin{equation}
\mathbb{Q}^{(p^{+}p^{-})}=\left( 
\begin{array}{ll}
0 & \text{$\mathbb{Q}$}^{(p+)} \\ 
\text{$\mathbb{Q}$}^{(p-)} & 0%
\end{array}%
\right) ,  \tag{3.32}
\end{equation}%
which must be prescribed in conformity to the scheme (2.15), with either of
the occurrent $\mathbb{Q}^{(p\pm )}$-pieces being taken as the inverse of
the other. Equation (3.32) does not involve any restrictions, but its
constituents are by definition continuous operators that correspond to the
observables $Q^{(p\pm )}$. We have the associations%
\begin{equation}
<e_{(0)}^{(p\pm )}\mid \text{$\mathbb{Q}$}^{(p\pm )}=<e_{(1)}^{(p\mp )}\mid ,%
\text{ }<e_{(1)}^{(p\pm )}\mid \text{$\mathbb{Q}$}^{(p\pm )}=<e_{(0)}^{(p\mp
)}\mid ,  \tag{3.33}
\end{equation}%
which promptly produce the entries%
\begin{equation}
\mathbb{Q}_{\mu \nu }^{(p\pm )}{}=<e_{(\mu )}^{(p\pm )}\mid \text{$\mathbb{Q}
$}^{(p\pm )}\parallel e_{(\nu )}^{(p\mp )}>_{g^{\mp }}=\text{$\mathbb{Q}$}%
_{\mu }^{(p\pm )\lambda }{}g_{\lambda \nu }^{\mp }.  \tag{3.34}
\end{equation}%
The mutual-inverse property of $\mathbb{Q}^{(p\pm )}$ yields the equivalent
relations (see (2.32))%
\begin{equation}
\mathbb{Q}_{\mu }^{(p\pm )\lambda }{}\mathbb{Q}_{\lambda }^{(p\mp )\nu
}{}=\Delta _{\mu \lambda }^{\pm }\Delta _{\pm }^{\ast \lambda \nu }=\mathbb{Q%
}_{\mu \lambda }^{(p\pm )}{}g_{\mp }^{\ast \lambda \sigma }\mathbb{Q}%
_{\sigma \tau }^{(p\mp )}{}g_{\pm }^{\ast \tau \nu },  \tag{3.35}
\end{equation}%
and%
\begin{equation}
<e_{(\mu )}^{(p\pm )}\mid \text{$\mathbb{Q}$}^{(p\pm )}\text{$\mathbb{Q}$}%
^{(p\mp )}\parallel e_{(\nu )}^{(p\pm )}>_{g^{\pm }}=g_{\mu \nu }^{\pm }. 
\tag{3.36}
\end{equation}%
Hence, we can write the overall charge-conjugation representation%
\begin{equation}
(\text{$\mathbb{Q}$}_{\mu \nu }^{(p^{+}p^{-})})=%
\begin{pmatrix}
0 & 0 & 0 & -1 \\ 
0 & 0 & -1 & 0 \\ 
0 & +1 & 0 & 0 \\ 
+1 & 0 & 0 & 0%
\end{pmatrix}%
,  \tag{3.37}
\end{equation}%
which appropriately carries the reduced contributions%
\begin{equation}
(\text{$\mathbb{Q}$}_{\mu \nu }^{(p\pm )}{})=\left( 
\begin{array}{ll}
0 & \mp 1 \\ 
\mp 1 & 0%
\end{array}%
\right) .  \tag{3.38}
\end{equation}

If we account for the statements%
\begin{equation}
<\Phi ^{(p\pm )}\mid \text{$\mathbb{Q}$}^{(p\pm )}=<\Phi ^{(p\mp )}\mid
\Leftrightarrow \Phi _{(p\pm )}^{\lambda }\text{$\mathbb{Q}$}_{\lambda
}^{(p\pm )\mu }{}=\Phi _{(p\mp )}^{\mu },  \tag{3.39}
\end{equation}%
then the action of $\mathbb{Q}^{(p^{+}p^{-})}$ on $<\Phi ^{(p^{+}p^{-})}\mid 
$ shall have to satisfy%
\begin{equation}
<\Phi ^{(p^{+}p^{-})}\mid \text{$\mathbb{Q}$}^{(p^{+}p^{-})}\parallel \Phi
^{(p^{+}p^{-})}>_{g}=<\Phi ^{(p^{+}p^{-})}\mid \Phi ^{(p^{+}p^{-})}>_{g}. 
\tag{3.40}
\end{equation}%
It follows that, joining together (3.30)-(3.34) suitably, provides us with
the restricted entries%
\begin{equation}
<e_{(\mu )}^{(p\pm )}\mid Q^{(p\pm )}\text{$\mathbb{Q}$}^{(p\pm )}Q^{(p\mp )}%
\text{$\mathbb{Q}$}^{(p\mp )}\parallel e_{(\nu )}^{(p\pm )}>_{g^{\pm
}}=-q^{2}g_{\mu \nu }^{\pm }.  \tag{3.41}
\end{equation}%
We can therefore say that the non-vanishing pieces of the array%
\begin{equation}
q^{(p^{+}p^{-})}=\left( 
\begin{array}{ll}
Q^{(p+)}\text{$\mathbb{Q}$}^{(p+)}Q^{(p-)}\text{$\mathbb{Q}$}^{(p-)} & 0 \\ 
0 & Q^{(p-)}\text{$\mathbb{Q}$}^{(p-)}Q^{(p+)}\text{$\mathbb{Q}$}^{(p+)}%
\end{array}%
\right) ,  \tag{3.42}
\end{equation}%
carry an observable character.

Taking the $\bigstar $-conjugate of the prescriptions (2.15) interchanges
the operator actions of $A^{+-}$ and $A^{-+}$, whence this conjugation
somehow replaces $(p\pm )$ with $(p\mp )$ in the case of $\mathbb{Q}^{(p\pm
)}$. A glance at the operator%
\begin{equation}
\text{$\mathbb{Q}$}^{(p^{+}p^{-})\bigstar }=\left( 
\begin{array}{ll}
0 & \text{$\mathbb{Q}$}^{(p-)\bigstar } \\ 
\text{$\mathbb{Q}$}^{(p+)\bigstar } & 0%
\end{array}%
\right) ,  \tag{3.43}
\end{equation}%
thus tells us that the $\bigstar $-version of (3.33) can be achieved
formally from the coupled correspondences%
\begin{equation}
<e_{(\mu )}^{(p\mp )}\mid \text{$\mathbb{Q}$}^{(p\pm )\bigstar
}\leftrightarrow <e_{(\mu )}^{(p\pm )}\mid \text{$\mathbb{Q}$}^{(p\pm )}. 
\tag{3.44}
\end{equation}%
Hence, if we call for the relations%
\begin{equation}
<e_{(\mu )}^{(p\mp )}\mid \text{$\mathbb{Q}$}^{(p\pm )\bigstar }\parallel
e_{(\nu )}^{(p\pm )}>_{g^{\pm }}=<e_{(\mu )}^{(p\pm )}\mid \text{$\mathbb{Q}$%
}^{(p\pm )}\parallel e_{(\nu )}^{(p\mp )}>_{g^{\mp }},  \tag{3.45}
\end{equation}%
likewise recalling (3.34), we will conclude that%
\begin{equation}
\text{$\mathbb{Q}$}_{\mu }^{(p\pm )\lambda }{}g_{\lambda \nu }^{\mp }=\text{$%
\mathbb{Q}$}_{\mu }^{(p\pm )\bigstar \lambda }{}g_{\lambda \nu }^{\pm
}\Rightarrow \mathbb{Q}_{\mu }^{(p\pm )\nu }{}=-\mathbb{Q}_{\mu }^{(p\pm
)\bigstar \nu }{}.  \tag{3.46}
\end{equation}%
The pattern (3.45) displays the characteristic pseudo-antiHermiticity
property%
\begin{equation}
\mathbb{Q}^{(p\pm )\bigstar }=-\mathbb{Q}^{(p\mp )}=-(\mathbb{Q}^{(p\pm
)})^{-1},  \tag{3.47}
\end{equation}%
since the entries (3.35) yield $\mathbb{Q}$$_{\mu \nu }^{(p+)}{}=-\mathbb{Q}$%
$_{\mu \nu }^{(p-)}{}$. Accordingly, we have the computation%
\begin{eqnarray}
&<&e_{(\mu )}^{(p\pm )}\mid \text{$\mathbb{Q}$}^{(p\pm )}\mid \text{$\mathbb{%
Q}$}^{(p\pm )}\mid e_{(\nu )}^{(p\pm )}>_{g^{\pm }}  \notag \\
&=&<e_{(\mu )}^{(p\pm )}\mid \text{$\mathbb{Q}$}^{(p\pm )}\text{$\mathbb{Q}$}%
^{(p\pm )\bigstar }\parallel e_{(\nu )}^{(p\pm )}>_{g^{\pm }}  \notag \\
&=&-<e_{(\mu )}^{(p\pm )}\mid e_{(\nu )}^{(p\pm )}>_{g^{\pm }}=-g_{\mu \nu
}^{\pm }=g_{\mu \nu }^{\mp }.  \TCItag{3.48}
\end{eqnarray}

If instead of the amplitude correspondence of (3.39) the antilinear
relationship $\overline{\Phi _{(p\pm )}^{\lambda }}\mathbb{Q}$$_{\lambda
}^{(p\pm )\mu }{}=\Phi _{(p\mp )}^{\mu }$ had been chosen, then a
pseudo-antiunitarity property could also be ascribed to $\mathbb{Q}$$^{(p\pm
)}$ via (3.47). However, in contrast to Dirac's theory, the action of
complex conjugation does not play any significant role in our specification
of charge and energy values. In passing, we notice that the decomposition
(2.17) was prescribed so as to let its operator pieces act on the right of
elements of $\mathfrak{C}$. This prescription is therefore distinct from the
one used in Refs. [16, 17]. We had still adopted it upon arranging the
pieces for (3.45), and it will be utilized again in Section 6.

Whenever either an up-down spin description of fermions or a
vertical-horizontal polarization description of bosons is to be carried out
in the given frame, a copy of the reduced canonical basis must indeed be
considered as an appropriate computational device. Let $(p^{+},p^{-})$ be a
massive fermionic pair $(f^{+},f^{-})$. We have the restricted up-down
equations%
\begin{equation}
<e_{(0)}^{(f\pm )}\mid \Sigma ^{(f\pm )}=+\frac{1}{2}<e_{(0)}^{(f\pm )}\mid ,%
\text{ }<e_{(1)}^{(f\pm )}\mid \Sigma ^{(f\pm )}=-\frac{1}{2}<e_{(1)}^{(f\pm
)}\mid ,  \tag{3.49}
\end{equation}%
together with the spin operator%
\begin{equation}
\Sigma ^{(f^{+}f^{-})}=%
\begin{pmatrix}
\Sigma ^{(f+)} & 0 \\ 
0 & \Sigma ^{(f-)}%
\end{pmatrix}%
,  \tag{3.50}
\end{equation}%
and the overall spin matrices%
\begin{equation}
(\Sigma _{\mu \nu }^{(f^{+}f^{-})})=%
\begin{pmatrix}
+\frac{1}{2} & 0 & 0 & 0 \\ 
0 & -\frac{1}{2} & 0 & 0 \\ 
0 & 0 & -\frac{1}{2} & 0 \\ 
0 & 0 & 0 & +\frac{1}{2}%
\end{pmatrix}%
=(\Sigma _{(f^{+}f^{-})}^{\ast \mu \nu }).  \tag{3.51}
\end{equation}%
The adjoint restrictions $\Sigma ^{(f\pm )}$ and $\Sigma _{(f\pm )}^{\ast }$
are thus represented by the traceless reduced blocks%
\begin{equation}
(\Sigma _{\mu \nu }^{(f\pm )})=%
\begin{pmatrix}
\pm \frac{1}{2} & 0 \\ 
0 & \mp \frac{1}{2}%
\end{pmatrix}%
=(\Sigma _{(f\pm )}^{\ast \mu \nu }).  \tag{3.52}
\end{equation}

When $(p^{+},p^{-})$ is a bosonic pair $(b^{+},b^{-})$ of any rest mass, the
observable configurations that carry its vertical-horizontal polarization
degrees of freedom emerge from the restrictions $\Pi ^{(b\pm )}$ of a
polarization operator $\Pi ^{(b^{+}b^{-})}$, according to%
\begin{equation}
<e_{(0)}^{(b\pm )}\mid \Pi ^{(b\pm )}=(+1)<e_{(0)}^{(b\pm )}\mid ,\text{ }%
<e_{(1)}^{(b\pm )}\mid \Pi ^{(b\pm )}=(-1)<e_{(1)}^{(b\pm )}\mid . 
\tag{3.53}
\end{equation}%
The corresponding restricted representations are written as%
\begin{equation}
(\Pi _{\mu \nu }^{(b\pm )})=%
\begin{pmatrix}
\pm 1 & 0 \\ 
0 & \mp 1%
\end{pmatrix}%
=(\Pi _{(b\pm )}^{\ast \mu \nu }).  \tag{3.54}
\end{equation}%
Amplitudes of bosonic states may be involved in the preparation of an
experimental setup for measuring elliptical, circular and linear
polarizations locally. The assignment between basis elements and
vertical-horizontal modes bears arbitrariness.

It will be made clear in Section 7 that every degenerate spectrum is $%
\mathcal{P}_{+}^{\uparrow }$-invariant whereas non-degenerate ones may be of
either behavioural type. By definition, the operator (3.32) requires the
particles $p^{+}$ and $p^{-}$ to carry reversed spin-polarization values and
non-invariant opposite-value total energies. Hence, supposing that the state
(3.2) is prepared such that either $p^{+}$ or $p^{-}$ carries a positive
total energy $E$, we deduce that the full spectra of the energy restrictions 
$H^{(p\pm )}$ for $(p^{+},p^{-})$ have to be constructed from non-degenerate
configurations like%
\begin{equation}
\begin{array}{c}
<e_{(0)}^{(p\pm )}\mid H_{I}^{(p\pm )}=\pm E<e_{(0)}^{(p\pm )}\mid \\ 
<e_{(1)}^{(p\pm )}\mid H_{I}^{(p\pm )}=\mp E<e_{(1)}^{(p\pm )}\mid \\ 
<e_{(0)}^{(p\pm )}\mid H_{II}^{(p\pm )}=\mp E<e_{(0)}^{(p\pm )}\mid \\ 
<e_{(1)}^{(p\pm )}\mid H_{II}^{(p\pm )}=\pm E<e_{(1)}^{(p\pm )}\mid .%
\end{array}
\tag{3.55a}
\end{equation}%
Each of the restrictions $H^{(p\pm )}$ has, in effect, to be made out as two
contributions, in accordance with the prescriptions%
\begin{equation}
H^{(p\pm )}=H_{I}^{(p\pm )}-H_{II}^{(p\pm )},  \tag{3.55b}
\end{equation}%
and%
\begin{equation}
(H_{I\mu \nu }^{(p\pm )})=\left( 
\begin{array}{ll}
E & 0 \\ 
0 & -E%
\end{array}%
\right) ,\text{ }(H_{II\mu \nu }^{(p\pm )})=\left( 
\begin{array}{ll}
-E & 0 \\ 
0 & E%
\end{array}%
\right) .  \tag{3.55c}
\end{equation}%
Suppressing the $p$-label for a moment, we see that the eigenvalues involved
in (3.55) enter the charge-conjugation relationships%
\begin{equation}
\begin{array}{c}
<e_{(0)}^{\pm }\mid H_{I}^{\pm }=\pm E<e_{(0)}^{\pm }\mid {\small %
\leftrightarrow }<e_{(0)}^{\pm }\mid \text{$\mathbb{Q}$}^{\pm }H_{II}^{\mp
}=\mp E<e_{(1)}^{\mp }\mid \\ 
<e_{(1)}^{\pm }\mid H_{I}^{\pm }=\mp E<e_{(1)}^{\pm }\mid {\small %
\leftrightarrow }<e_{(1)}^{\pm }\mid \text{$\mathbb{Q}$}^{\pm }H_{II}^{\mp
}=\pm E<e_{(0)}^{\mp }\mid .%
\end{array}
\tag{3.56}
\end{equation}%
They also occur in the subsidiary associations%
\begin{equation}
\begin{array}{c}
<e_{(0)}^{+}\mid H_{I}^{+}=+E<e_{(0)}^{+}\mid {\small \leftrightarrow }%
<e_{(0)}^{+}\mid v^{+}H_{I}^{+}=-E<e_{(1)}^{+}\mid \\ 
<e_{(1)}^{-}\mid H_{I}^{-}=+E<e_{(1)}^{-}\mid {\small \leftrightarrow }%
<e_{(1)}^{-}\mid v^{-}H_{I}^{-}=-E<e_{(0)}^{-}\mid ,%
\end{array}
\tag{3.57a}
\end{equation}%
and%
\begin{equation}
\begin{array}{c}
<e_{(1)}^{+}\mid H_{II}^{+}=+E<e_{(1)}^{+}\mid {\small \leftrightarrow }%
<e_{(1)}^{+}\mid v^{+}H_{II}^{+}=-E<e_{(0)}^{+}\mid \\ 
<e_{(0)}^{-}\mid H_{II}^{-}=+E<e_{(0)}^{-}\mid {\small \leftrightarrow }%
<e_{(0)}^{-}\mid v^{-}H_{II}^{-}=-E<e_{(1)}^{-}\mid ,%
\end{array}
\tag{3.57b}
\end{equation}%
with the definitions%
\begin{equation}
<e_{(0)}^{(p\pm )}\mid v^{(p\pm )}\doteqdot <e_{(1)}^{(p\pm )}\mid ,\text{ }%
<e_{(1)}^{(p\pm )}\mid v^{(p\pm )}\doteqdot <e_{(0)}^{(p\pm )}\mid , 
\tag{3.57c}
\end{equation}%
which prescribe in the given frame what we call the virtual-particle
restrictions for $(p^{+},p^{-})$ along with the unstarred basis states for
the respective virtual particles.

The virtual-particle operators for charged fermionic or bosonic pairs,
amount to pseudo-Hermitian involutions that interchange the signs of the
relevant spin-polarization values and energies without affecting any charges
at all. They should therefore take over the role of charge conjugations
whenever particles and antiparticles are considered individually. Fitting
them together with charge conjugations makes it possible to accomplish
locally the non-degenerate completeness of structures like (3.55) from
disjoint unreduced representations of the type%
\begin{equation}
(v_{\mu \nu }^{(p+)})=%
\begin{pmatrix}
0 & 1 & 0 & 0 \\ 
1 & 0 & 0 & 0 \\ 
0 & 0 & 0 & 0 \\ 
0 & 0 & 0 & 0%
\end{pmatrix}%
,\text{ }(v_{\mu \nu }^{(p-)})=%
\begin{pmatrix}
0 & 0 & 0 & 0 \\ 
0 & 0 & 0 & 0 \\ 
0 & 0 & 0 & -1 \\ 
0 & 0 & -1 & 0%
\end{pmatrix}%
.  \tag{3.58}
\end{equation}%
We thus have the charge-conjugation and virtual-particle correspondences
borne by the spin-polarization-energy schemes%
\begin{equation}
\begin{array}{ccc}
\text{$\mathbb{Q}^{(f+)}$}:\pm E\uparrow \downarrow \text{for }f^{+} & 
{\LARGE \mapsto } & \mp E\downarrow \uparrow \text{for }f^{-} \\ 
\text{$\mathbb{Q}^{(b+)}$}:\pm E\updownarrow \leftrightarrow \text{for }b^{+}
& {\LARGE \mapsto } & \mp E\leftrightarrow \updownarrow \text{for }b^{-},%
\end{array}
\tag{3.59a}
\end{equation}%
and%
\begin{equation}
\begin{array}{ccc}
v^{(f\pm )}:\pm E\uparrow \downarrow \text{for }f^{\pm } & {\LARGE \mapsto }
& \mp E\downarrow \uparrow \text{for }f^{\pm } \\ 
v^{(b\pm )}:\pm E\updownarrow \leftrightarrow \text{for }b^{\pm } & {\LARGE %
\mapsto } & \mp E\leftrightarrow \updownarrow \text{for }b^{\pm },%
\end{array}
\tag{3.59b}
\end{equation}%
together with the inverse version of (3.59a). The symbols $\uparrow
\downarrow $ and $\updownarrow \leftrightarrow $ have been used to denote
pictorially the up-down and vertical-horizontal states carried by (3.49) and
(3.53).

There is a compelling reason for choosing the reversed spin-polarization
values fixed up by (3.33) and (3.57c), which appears to be related to the
disconnectedness of the representations (3.58). This point will be
reconsidered at greater length in Section 8. We emphasize once again that
energy restrictions must be defined as formal operators which are not
identified with any Hamiltonians. If $(p^{+},p^{-})$ were chosen to carry
vanishing $(\pm q)$-charge values, the scheme (3.59a) would become
meaningless. In the Dirac context, massive particles and antiparticles are
oftenly taken to carry opposite-value energies because they could otherwise
propagate outside light cones, but the actions of parity and time-reversal
operators may in any case yield a spacetime-direction commonness (see Ref.
[28]).

The classes of unitary operators on $\mathfrak{C}^{\pm }$ and $\mathfrak{C}%
_{\pm }^{\ast }$ that may be used for changing locally the description of
the degrees of freedom of $(p^{+},p^{-})$ particularly supply all the
realizable descriptions of spins and polarizations other than those of the
type afforded by (3.49) and (3.53). In fact, some of the most interesting
manipulations produce a formal definition of helicities for fermions and
bosons as well as an interchange between up-down and vertical-horizontal
attributes. As for the case of (3.13), the operators that yield such
alternative configurations carry an intrinsically local character whence
their matrix representations do not belong to the intersection (3.12). The
main procedure for carrying out any modification picks out some local
unitary restrictions $\mathfrak{u}^{\pm }$, and implements prescriptions of
the form%
\begin{equation}
<\mathfrak{e}_{(\mu )}^{(p\pm )}\mid =<e_{(\mu )}^{(p\pm )}\mid \mathfrak{u}%
^{\pm },\text{ }<\mathfrak{e}_{(p\pm )}^{\ast (\mu )}\mid =<e_{(p\pm
)}^{\ast (\mu )}\mid \mathfrak{u}_{\pm }^{\ast },  \tag{3.60}
\end{equation}%
which imply that%
\begin{equation}
<<\mathfrak{e}_{(\mu )}^{(p\pm )}\mid \mathfrak{e}_{(\nu )}^{(p\pm
)}>>=\Delta _{\mu \nu }^{\pm },\text{ }<<\mathfrak{e}_{(p\pm )}^{\ast (\mu
)}\mid \mathfrak{e}_{(p\pm )}^{\ast (\nu )}>>=\Delta _{\pm }^{\ast \mu \nu }.
\tag{3.61}
\end{equation}

Under the changes (3.60), the amplitudes of the $\Phi $-state may be left
invariant together with the normalization condition (3.3) such that the
state itself could transform as%
\begin{equation}
<\Phi ^{(p\pm )}\mid \mapsto <\varphi ^{(p\pm )}\mid =\varphi _{(p\pm
)}^{\mu }<\mathfrak{e}_{(\mu )}^{(p\pm )}\mid =\Phi _{(p\pm )}^{\lambda }%
\mathfrak{u}_{\lambda }^{\pm }{}^{\mu }<e_{(\mu )}^{(p\pm )}\mid . 
\tag{3.62}
\end{equation}%
We have the configuration%
\begin{equation}
<\mathfrak{e}_{(\mu )}^{(p\pm )}\mid a^{(p\pm )}\parallel \mathfrak{e}_{(\nu
)}^{(p\pm )}>_{\mathfrak{G}^{(u\pm )}}=<e_{(\mu )}^{(p\pm )}\mid A^{(p\pm
)}\parallel e_{(\nu )}^{(p\pm )}>_{g^{\pm }},  \tag{3.63}
\end{equation}%
which shows that for the eigenvalues of $A^{(p\pm )}$ to be preserved under
the implementation of (3.60), we should take account of the restricted
relationships%
\begin{equation}
A^{(p\pm )}=\mathfrak{u}^{\pm }a^{(p\pm )}\mathfrak{u}^{\pm \dag },\text{ }%
A_{(p\pm )}^{\ast }=\mathfrak{u}_{\pm }^{\ast }a_{(p\pm )}^{\ast }\mathfrak{u%
}_{\pm }^{\ast \dag },  \tag{3.64a}
\end{equation}%
along with the Gram restrictions%
\begin{equation}
\mathfrak{G}^{(u\pm )}=\mathfrak{u}^{\pm \dag }g^{\pm }\mathfrak{u}^{\pm },%
\text{ }\mathfrak{G}_{(u\pm )}^{\ast }=\mathfrak{u}_{\pm }^{\ast \dag
}g_{\pm }^{\ast }\mathfrak{u}_{\pm }^{\ast }.  \tag{3.64b}
\end{equation}%
Hence, by setting%
\begin{equation}
a_{\mu \nu }^{(p\pm )}=<\mathfrak{e}_{(\mu )}^{(p\pm )}\mid a^{(p\pm
)}\parallel \mathfrak{e}_{(\nu )}^{(p\pm )}>_{\mathfrak{G}^{(u\pm )}}, 
\tag{3.65}
\end{equation}%
we get the local spectral equalities%
\begin{equation}
a_{\mu \nu }^{(p\pm )}=A_{\mu \nu }^{(p\pm )},\text{ }a_{(p\pm )}^{\ast \mu
\nu }=A_{(p\pm )}^{\ast \mu \nu },  \tag{3.66}
\end{equation}%
together with\footnote{%
Equations (3.37) and (3.58) must be preserved under the changes (3.60). In
Section 8, we will elaborate upon the corresponding operator behaviours.}%
\begin{equation}
<\mathfrak{e}_{(\mu )}^{(p\pm )}\mid \mathfrak{e}_{(\nu )}^{(p\pm )}>_{%
\mathfrak{G}^{(u\pm )}}=<e_{(\mu )}^{(p\pm )}\mid e_{(\nu )}^{(p\pm
)}>_{g^{\pm }}.  \tag{3.67}
\end{equation}

When (3.60) and (3.63) are implemented, both the traces (3.20) and the
product (3.25) get preserved along with the completeness relation (2.40) and
the definiteness of (2.28). Additionally, any relationships like those of
(3.64) ensure the preservation of the pseudo Hermiticity of the manipulated
observables, as can be seen by utilizing the operator prescriptions%
\begin{equation}
A^{(p\pm )\dag }=g^{\pm }A^{(p\pm )\bigstar }g^{\pm },\text{ }a^{(p\pm )\dag
}=\mathfrak{G}^{(u\pm )}a^{(p\pm )\#}\mathfrak{G}^{(u\pm )},  \tag{3.68}
\end{equation}%
where the symbol $\#$ stands for the operation of pseudo-Hermitian
conjugation with respect to the $\mathfrak{GG}^{\ast }$-inner products. The
crucial point as regards the latter property is associated to the
applicability of the local relations%
\begin{equation}
A^{(p\pm )\bigstar }=\mathfrak{u}^{\pm }a^{(p\pm )\#}\mathfrak{u}^{\pm \dag
},\text{ }A_{(p\pm )}^{\ast \bigstar }=\mathfrak{u}_{\pm }^{\ast }a_{(p\pm
)}^{\ast \#}\mathfrak{u}_{\pm }^{\ast \dag },  \tag{3.69}
\end{equation}%
which assure that the requirements%
\begin{equation}
A^{(p\pm )}=A^{(p\pm )\bigstar },\text{ }A_{(p\pm )}^{\ast }=A_{(p\pm
)}^{\ast \bigstar },  \tag{3.70a}
\end{equation}%
and 
\begin{equation}
a^{(p\pm )}=a^{(p\pm )\#},\text{ }a_{(p\pm )}^{\ast }=a_{(p\pm )}^{\ast \#},
\tag{3.70b}
\end{equation}%
are mutually satisfied. The relevant expectation values are therefore
subject to 
\begin{equation}
<\Phi ^{(p\pm )}\mid A^{(p\pm )}\parallel \Phi ^{(p\pm )}>_{g^{\pm
}}=<\varphi ^{(p\pm )}\mid a^{(p\pm )}\parallel \varphi ^{(p\pm )}>_{%
\mathfrak{G}^{(u\pm )}},  \tag{3.71a}
\end{equation}%
whence%
\begin{equation}
<\Phi ^{(p\pm )}\mid \Phi ^{(p\pm )}>_{g^{\pm }}=<\varphi ^{(p\pm )}\mid
\varphi ^{(p\pm )}>_{\mathfrak{G}^{(u\pm )}}.  \tag{3.71b}
\end{equation}%
For any observable restrictions that obey the prescriptions (3.64a), we also
have the commutator property%
\begin{equation}
\lbrack A^{(p\pm )},B^{(p\pm )}]=\mathfrak{u}^{\pm }[a^{(p\pm )},b^{(p\pm )}]%
\mathfrak{u}^{\pm \dag }.  \tag{3.72}
\end{equation}

The $\Phi $-state could arbitrarily have been chosen to behave invariantly
under the changes (3.60). Making this choice would still preserve the
strongly required definite normalization condition whilst changing the
individual probabilities (3.10), and the invariant behaviour exhibited by
(3.71) would likewise cease holding.

It has become manifest that it is either of the basis transformations (3.60)
which allows us to change the $\Sigma \Pi $-descriptions. The helicity
restrictions $h^{(p\pm )}$ for $(p^{+},p^{-})$ are defined together with a
selection of $\mathfrak{u}^{\pm }$ that yields suitable states. For the case
of any massive (massless) particle or antiparticle, the definition of a
helicity operator takes up implicitly some timelike (null) direction in
spacetime as the pertinent locally specified direction of motion.\footnote{%
Defining helicity spectra for massless fermions demands at least in the
first instance a modification of the matrices (3.52).} Hence, changing local
spacetime directions, produces the requirement for selecting other $%
\mathfrak{u}$-operators. Typically, we have the bases%
\begin{equation}
<\mathfrak{h}_{(\mu )}^{(p\pm )}\mid =<e_{(\mu )}^{(p\pm )}\mid \text{%
\textsc{H}}^{(p\pm )},\text{ }<\mathfrak{h}_{(p\pm )}^{\ast (\mu )}\mid
=<e_{(p\pm )}^{\ast (\mu )}\mid \text{\textsc{H}}_{(p\pm )}^{\ast }, 
\tag{3.73a}
\end{equation}%
which supply us with the left-right states%
\begin{equation}
<\mathfrak{h}_{(0)}^{(p\pm )}\mid \doteqdot <L^{(p\pm )}\mid ,\text{ }<%
\mathfrak{h}_{(1)}^{(p\pm )}\mid \doteqdot <R^{(p\pm )}\mid .  \tag{3.73b}
\end{equation}%
In the fermionic case of $(f^{+},f^{-})$, we then obtain the helicity
spectrum%
\begin{equation}
h_{\mu \nu }^{(f\pm )}=<\mathfrak{h}_{(\mu )}^{(f\pm )}\mid h^{(f\pm
)}\parallel \mathfrak{h}_{(\nu )}^{(f\pm )}>_{\mathfrak{H}^{(f\pm )}}=\Sigma
_{\mu \nu }^{(f\pm )},  \tag{3.74}
\end{equation}%
with the prescriptions (3.64) thus yielding the operators%
\begin{equation}
h^{(f\pm )}=\text{\textsc{H}}^{(f\pm )\dag }\Sigma ^{(f\pm )}\text{\textsc{H}%
}^{(f\pm )},\text{ }\mathfrak{H}^{(f\pm )}=\text{\textsc{H}}^{(f\pm )\dag
}g^{\pm }\text{\textsc{H}}^{(f\pm )}.  \tag{3.75}
\end{equation}%
Similarly, the left-right description of the bosonic pair $(b^{+},b^{-})$ is
provided by%
\begin{equation}
h_{\mu \nu }^{(b\pm )}=<\mathfrak{h}_{(\mu )}^{(b\pm )}\mid h^{(b\pm
)}\parallel \mathfrak{h}_{(\nu )}^{(b\pm )}>_{\mathfrak{H}^{(b\pm )}}=\Pi
_{\mu \nu }^{(b\pm )},  \tag{3.76}
\end{equation}%
with%
\begin{equation}
h^{(b\pm )}=\text{\textsc{H}}^{(b\pm )\dag }\Pi ^{(b\pm )}\text{\textsc{H}}%
^{(b\pm )},\text{ }\mathfrak{H}^{(b\pm )}=\text{\textsc{H}}^{(b\pm )\dag
}g^{\pm }\text{\textsc{H}}^{(b\pm )}.  \tag{3.77}
\end{equation}%
It should be evident that the restrictions given by (3.57c) reverse
helicities as well.

An interchange between the dynamical characters of the $\Sigma \Pi $%
-descriptions may be attained from the configurations%
\begin{equation}
<\mathfrak{p}_{(\mu )}^{(f\pm )}\mid =<e_{(\mu )}^{(f\pm )}\mid \text{%
\textsc{P}}^{(f\pm )},\text{ }\mathfrak{P}^{(f\pm )}=\text{\textsc{P}}%
^{(f\pm )\dag }g^{\pm }\text{\textsc{P}}^{(f\pm )},  \tag{3.78}
\end{equation}%
and%
\begin{equation}
<\mathfrak{z}_{(\mu )}^{(b\pm )}\mid =<e_{(\mu )}^{(b\pm )}\mid \text{%
\textsc{Z}}^{(b\pm )},\text{ }\mathfrak{Z}^{(b\pm )}=\text{\textsc{Z}}%
^{(b\pm )\dag }g^{\pm }\text{\textsc{Z}}^{(b\pm )},  \tag{3.79}
\end{equation}%
which correspondingly yield the vertical-horizontal fermionic bases%
\begin{equation}
<\mathfrak{p}_{(0)}^{(f\pm )}\mid =<V^{(f\pm )}\mid ,\text{ }<\mathfrak{p}%
_{(1)}^{(f\pm )}\mid =<H^{(f\pm )}\mid ,  \tag{3.80}
\end{equation}%
together with the up-down bosonic ones%
\begin{equation}
<\mathfrak{z}_{(0)}^{(b\pm )}\mid =<U^{(b\pm )}\mid ,\text{ }<\mathfrak{z}%
_{(1)}^{(b\pm )}\mid =<D^{(b\pm )}\mid .  \tag{3.81}
\end{equation}%
We are thus led to the spectra%
\begin{equation}
\Pi _{\mu \nu }^{(f\pm )}=<\mathfrak{p}_{(\mu )}^{(f\pm )}\mid \Pi ^{(f\pm
)}\parallel \mathfrak{p}_{(\nu )}^{(f\pm )}>_{\mathfrak{P}^{(f\pm )}}, 
\tag{3.82}
\end{equation}%
and%
\begin{equation}
\Sigma _{\mu \nu }^{(b\pm )}=<\mathfrak{z}_{(\mu )}^{(b\pm )}\mid \Sigma
^{(b\pm )}\parallel \mathfrak{z}_{(\nu )}^{(b\pm )}>_{\mathfrak{Z}^{(b\pm
)}},  \tag{3.83}
\end{equation}%
along with the observable relationships%
\begin{equation}
\Sigma ^{(f\pm )}=\text{\textsc{P}}^{(f\pm )}\Pi ^{(f\pm )}\text{\textsc{P}}%
^{(f\pm )\dag },\text{ }\Pi ^{(b\pm )}=\text{\textsc{Z}}^{(b\pm )}\Sigma
^{(b\pm )}\text{\textsc{Z}}^{(b\pm )\dag }.  \tag{3.84}
\end{equation}

\section{Density operators, entropies and composite states}

To any locally prepared state for some particle or antiparticle, we ascribe
a density operator which affords us an alternative methodology as well as a $%
\mathcal{P}_{+}^{\uparrow }$-invariant definition of entropies. For the pair 
$(p^{+},p^{-})$, with the state (3.2), we express the respective densities
as the pseudo-Hermitian restrictions%
\begin{equation}
\rho ^{(\Phi \pm )}=g^{\pm }\mid \Phi ^{(p\pm )}><\Phi ^{(p\pm )}\mid =\mid
\Phi ^{(p\pm )}><\Phi ^{(p\pm )}\mid g^{\pm },  \tag{4.1}
\end{equation}%
with the property (2.29) obviously enabling us to relax the double-bar
convention. Hence, letting $\rho ^{(\Phi \pm )}$ act on its reduced states
adequately and making use of (3.3), yields the eigenvalue equations%
\begin{equation}
<\Phi ^{(p\pm )}\mid \rho ^{(\Phi \pm )}=(+1)<\Phi ^{(p\pm )}\mid , 
\tag{4.2a}
\end{equation}%
and%
\begin{equation}
\rho ^{(\Phi \pm )}\mid \Phi ^{(p\pm )}>=\mid \Phi ^{(p\pm )}>(+1), 
\tag{4.2b}
\end{equation}%
together with the values%
\begin{equation}
<\Phi ^{(p\pm )}\mid \rho ^{(\Phi \pm )}\parallel \Phi ^{(p\pm )}>_{g^{\pm
}}=\pm 1.  \tag{4.3}
\end{equation}%
The corresponding representative matrix is unambiguously set as\footnote{%
The off-diagonal entries of (4.4) define typical \textit{coherences} which
may describe an interference process.}%
\begin{equation}
(\rho _{\mu \nu }^{(\Phi \pm )})=%
\begin{pmatrix}
\pm \mid \Phi _{(p\pm )}^{0}\mid ^{2} & \pm \Phi _{(p\pm )}^{1}\overline{%
\Phi _{(p\pm )}^{0}} \\ 
\pm \Phi _{(p\pm )}^{0}\overline{\Phi _{(p\pm )}^{1}} & \pm \mid \Phi
_{(p\pm )}^{1}\mid ^{2}%
\end{pmatrix}%
,  \tag{4.4}
\end{equation}%
which equals the one that results from the implementation of the replacement%
\begin{equation}
\Phi _{(p\pm )}^{\mu }\leftrightarrow \Phi _{\mu }^{(p\pm )}.  \tag{4.5}
\end{equation}

We should notice that%
\begin{equation}
g^{\pm }\rho ^{(\Phi \pm )}=\rho ^{(\Phi \pm )}g^{\pm },\text{ }(\rho
^{(\Phi \pm )})^{2}=\rho ^{(\Phi \pm )},  \tag{4.6}
\end{equation}%
whence, allowing for (3.3) once more, leads to the positive-definite traces%
\begin{equation}
\text{Tr }\rho ^{(\Phi \pm )}=+1=\text{Tr }\rho _{(\Phi \pm )}^{\ast }, 
\tag{4.7}
\end{equation}%
whilst the value (3.21) turns out to be reexpressed as%
\begin{equation}
<A^{(p\pm )}>_{\Phi ^{\pm }}=\text{Tr }(g^{\pm }\rho ^{(\Phi \pm )}A^{(p\pm
)}).  \tag{4.8}
\end{equation}%
Dropping the Gram operator from the right-hand side of (4.8), produces the
(uninteresting) Hilbert value%
\begin{equation}
\text{Tr }(\rho ^{(\Phi \pm )}A^{(p\pm )})=<<\Phi ^{(p\pm )}\mid A^{(p\pm
)}\parallel \Phi ^{(p\pm )}>>.  \tag{4.9}
\end{equation}%
It becomes clear that the applicability of the law (3.13) induces the
occurrence of the local evolution equation%
\begin{equation}
g^{\pm }\rho ^{(\Phi \pm )}(\tau )=\mathfrak{U}^{\pm \bigstar }{}(\tau ,\tau
_{0})g^{\pm }\rho ^{(\Phi \pm )}(\tau _{0})\mathfrak{U}^{\pm }{}(\tau ,\tau
_{0}),  \tag{4.10}
\end{equation}%
which is invariant under the $\bigstar $-conjugation because of the
commutativity property of (4.6).

Any local description that takes up the prescriptions (3.60) and (3.64) may
involve decomposable density restrictions of the form%
\begin{equation}
\rho ^{(\varphi \pm )}=\frac{1}{2}\mid \mathfrak{e}_{(\mu )}^{(p\pm )}>%
\mathfrak{G}_{\pm }^{\ast \mu \nu }<\mathfrak{e}_{(\nu )}^{(p\pm )}\mid . 
\tag{4.11}
\end{equation}%
The density (4.11) thus satisfies the equation%
\begin{equation}
<\mathfrak{e}_{(\mu )}^{(p\pm )}\mid \rho ^{(\varphi \pm )}=+\frac{1}{2}<%
\mathfrak{e}_{(\mu )}^{(p\pm )}\mid ,  \tag{4.12}
\end{equation}%
which gives the entries%
\begin{equation}
<\mathfrak{e}_{(\mu )}^{(p\pm )}\mid \rho ^{(\varphi \pm )}\parallel 
\mathfrak{e}_{(\nu )}^{(p\pm )}>_{\mathfrak{G}\pm }=\frac{1}{2}\mathfrak{G}%
_{\mu \nu }^{\pm },  \tag{4.13}
\end{equation}%
along with the traces%
\begin{equation}
\text{Tr }\rho ^{(\varphi \pm )}=1=\text{Tr }\rho _{(\varphi \pm )}^{\ast }.
\tag{4.14}
\end{equation}%
It follows that, in the case of any canonical descriptions, we can implement
decompositions like%
\begin{equation}
\rho ^{(p\pm )}=\frac{1}{2}\mid e_{(\mu )}^{(p\pm )}>g_{\pm }^{\ast \mu \nu
}<e_{(\nu )}^{(p\pm )}\mid ,  \tag{4.15}
\end{equation}%
which produce representations of the type%
\begin{equation}
(\rho _{\mu \nu }^{(p\pm )})=%
\begin{pmatrix}
\pm \frac{1}{2} & 0 \\ 
0 & \pm \frac{1}{2}%
\end{pmatrix}%
.  \tag{4.16}
\end{equation}%
The matrix (4.16) evidently equals the one that represents (4.11).

Every state of a particle or antiparticle carries an intrinsic entropy which
is formally expressed and evaluated in the traditional manner (see Refs. [6,
29]). Hence, all the standard concavity and subadditivity entropic
properties are satisfied in the case of any of the densities we have just
built up. For the state (3.2), we thus have the expression%
\begin{equation}
S(\rho ^{(\Phi \pm )})=-\text{Tr }(\rho ^{(\Phi \pm )}\log _{2}\rho ^{(\Phi
\pm )}),  \tag{4.17}
\end{equation}%
which, when combined with (4.2), establishes that any single pure state for
a particle or antiparticle carries a vanishing entropy. Of course, the
density (4.11) yields the value%
\begin{equation}
S(\rho ^{(\varphi \pm )})=\log _{2}2,  \tag{4.18}
\end{equation}%
which coincides with that of $S(\rho ^{(p\pm )})$.

As was said in Section 1, any states of composite systems made out of
non-interacting particles and antiparticles amount to the tensor product of
the individual states ascribed to the physical constituents involved.%
\footnote{%
The description of composite systems need not involve pairs of
particle-antiparticle companions. Weighted sums (mixtures) of states will
not be taken into consideration here.} The values of composite amplitudes
are supplied by Kronecker products, and always bear an $SU(2,2)$-tensor
character. Thus, by adopting the notation%
\begin{equation}
<\Psi ^{(k\pm )}\mid =\Psi _{(k\pm )}^{\mu }<e_{(\mu )}^{(k\pm )}\mid ,\text{
}<\Psi _{(k\pm )}^{\ast }\mid =\Psi _{\mu }^{(k\pm )}<e_{(k\pm )}^{\ast (\mu
)}\mid ,  \tag{4.19}
\end{equation}%
where the $(k\pm )$-label refers to the individual states being composed as
well as to the respective adjoint copies of the canonical bases, we write
the configuration%
\begin{equation}
<\Psi ^{(N^{+}N^{-})}\mid =(\overset{N^{+}}{\underset{k=1}{\otimes }}<\Psi
^{(k+)}\mid )\otimes (\overset{N^{-}}{\underset{k=1}{\otimes }}<\Psi
^{(k-)}\mid ),  \tag{4.20}
\end{equation}%
with $N^{+}$ and $N^{-}$ standing for the numbers of particles and
antiparticles that constitute some composite system $\mathcal{S}%
_{N^{+}N^{-}} $, and the labels $k$ running independently of one another. It
is obvious that%
\begin{equation}
<\Psi ^{(N^{+}N^{-})}\mid \in (\overset{N^{+}}{\underset{k=1}{\otimes }}%
\mathfrak{C}^{(k+)})\otimes (\overset{N^{-}}{\underset{k=1}{\otimes }}%
\mathfrak{C}^{(k-)}),  \tag{4.21}
\end{equation}%
where $\mathfrak{C}^{(k\pm )}$ is spanned by the unstarred reduced basis of
(4.19).

The amplitude of the state (4.20) reads%
\begin{equation}
C_{(N^{+}N^{-})}^{\mu ...\nu \lambda ...\sigma }=\Psi _{(1+)}^{\mu }...\Psi
_{(N+)}^{\nu }\Psi _{(1-)}^{\lambda }...\Psi _{(N-)}^{\sigma }.  \tag{4.22}
\end{equation}%
It accordingly carries $(N^{+}+N^{-})$ indices, whence the dimension of
either of the adjoint spaces of composite states for $\mathcal{S}%
_{N^{+}N^{-}}$ equals $2^{N^{+}+N^{-}}$. Composite states may form sparse
subsets of product spaces such that they do not generally admit separability
or index symmetries. Writing down composite states requires making an
arbitrary choice of factor ordering without the necessity for keeping track
afterwards of the signs associated to eventually occurrent permutations of
states for indistinguishable fermions (see Section 8).

The Gram operators that define the $gg^{\ast }$-inner products on the spaces
of adjoint states for $\mathcal{S}_{N^{+}N^{-}}$, are prescribed as the
tensor juxtaposition of suitable numbers of copies of $g^{\pm }$ and $g_{\pm
}^{\ast }$. For instance,%
\begin{equation}
g^{(N^{+}N^{-})}=g^{[N^{+}]}\otimes g^{[N^{-}]},\text{ }g^{[N^{\pm
}]}\doteqdot \overset{N^{\pm }}{\underset{k=1}{\otimes }}g^{(k\pm )}, 
\tag{4.23}
\end{equation}%
with $g^{(k\pm )}$ thus operating on $\mathfrak{C}^{(k\pm )}$. For the
product-state pieces of (4.20), we have the formal pattern%
\begin{equation}
\overset{N^{\pm }}{\underset{k=1}{\otimes }}<\Psi ^{(k\pm )}\mid =\Psi
_{(1\pm )}^{\mu }...\Psi _{(N\pm )}^{\nu }<e_{(\mu )}^{(1\pm )}\mid \otimes
...\otimes <e_{(\nu )}^{(N\pm )}\mid ,  \tag{4.24}
\end{equation}%
whence implementing the shorthand notation%
\begin{equation}
<e_{(\mu )}^{(1\pm )}\mid \otimes ...\otimes <e_{(\nu )}^{(N\pm )}\mid
=<e_{(\mu )}^{(1\pm )}...e_{(\nu )}^{(N\pm )}\mid ,  \tag{4.25}
\end{equation}%
together with the relations (2.37), we get the components%
\begin{equation}
\Psi _{\mu }^{(1\pm )}...\Psi _{\nu }^{(N\pm )}=\overset{N^{\pm }}{\underset{%
k=1}{\otimes }}<\Psi ^{(k\pm )}\mid e_{(\mu )}^{(1\pm )}...e_{(\nu )}^{(N\pm
)}>_{g^{[N^{\pm }]}}.  \tag{4.26}
\end{equation}

A prototypical dynamical variable for $\mathcal{S}_{N^{+}N^{-}}$ is
expressed as 
\begin{equation}
A^{(N^{+}N^{-})}=(\overset{N^{+}}{\underset{k=1}{\otimes }}A^{(k+)})\otimes (%
\overset{N^{-}}{\underset{k=1}{\otimes }}A^{(k-)}),  \tag{4.27}
\end{equation}%
where $A^{(k\pm )}$ denotes the pertinent restriction for the $(k\pm )$%
-constituent of $\mathcal{S}_{N^{+}N^{-}}$. The local spectral
representation of the operator (4.27) emerges as the Kronecker product of
the matrices whose entries are given by%
\begin{equation}
A_{\mu \nu }^{(k\pm )}=<e_{(\mu )}^{(k\pm )}\mid A^{(k\pm )}\parallel
e_{(\nu )}^{(k\pm )}>_{g^{(k\pm )}}.  \tag{4.28}
\end{equation}%
To the expectation value of $A^{(N^{+}N^{-})}$ in the state (4.20), we have
the contribution%
\begin{equation}
<\Psi ^{\lbrack N^{\pm }]}\mid A^{[N^{\pm }]}\parallel \Psi ^{\lbrack N^{\pm
}]}>_{g^{[N^{\pm }]}}=\overset{N^{\pm }}{\underset{k=1}{\Pi }}<A^{(k\pm
)}>_{\Psi ^{(k\pm )}},  \tag{4.29}
\end{equation}%
with the square-bracket notation of (4.23) having been utilized. It is
possible to prepare composite states and perform locally one measurement at
a time or even several measurements at once. This ensures the physical
significance of (4.27)-(4.29).

Of considerable interest are the spectral matrices that come out when some
observable restriction is selected out of the product (4.27). In effect, by
selecting $A^{(j+)}$, with $1\leq j\leq N^{+}$, and writing the expression%
\begin{equation}
A^{(j^{+}N^{+}N^{-})}=I^{(1+)}\otimes ...\otimes A^{(j+)}\otimes ...\otimes
I^{(N+)}\otimes I^{(1-)}\otimes ...\otimes I^{(N-)},  \tag{4.30}
\end{equation}%
with any $I^{(r\pm )}$ amounting to the restriction on $\mathfrak{C}^{(r\pm
)}$ of the identity operator, we obtain the entry%
\begin{eqnarray}
&<&E_{(\alpha ...\mu ...\gamma \lambda ...\rho )}^{(j^{+}N^{+}N^{-})}\mid
A^{(j^{+}N^{+}N^{-})}\parallel E_{(\beta ...\nu ...\delta \sigma ...\tau
)}^{(j^{+}N^{+}N^{-})}>_{g^{(N^{+}N^{-})}}  \notag \\
&=&g_{\alpha \beta }^{(1+)}...A_{\mu \nu }^{(j+)}...g_{\gamma \delta
}^{(N+)}g_{\lambda \sigma }^{(1-)}...g_{\rho \tau }^{(N-)},  \TCItag{4.31}
\end{eqnarray}%
where we have used the outer-product notation%
\begin{equation}
<E_{(\alpha ...\mu ...\gamma \lambda ...\rho )}^{(j^{+}N^{+}N^{-})}\mid
=<e_{(\alpha )}^{(1+)}...e_{(\mu )}^{(j+)}...e_{(\gamma )}^{(N+)}e_{(\lambda
)}^{(1-)}...e_{(\rho )}^{(N-)}\mid .  \tag{4.32}
\end{equation}%
Now, by selecting $A^{(j-)}$, with $1\leq j\leq N^{-}$, and writing%
\begin{equation}
A^{(N^{+}j^{-}N^{-})}=I^{(1+)}\otimes ...\otimes I^{(N+)}\otimes
I^{(1-)}\otimes ...\otimes A^{(j-)}\otimes ...\otimes I^{(N-)},  \tag{4.33}
\end{equation}%
we similarly get%
\begin{eqnarray}
&<&E_{(\alpha ...\gamma \lambda ...\mu ...\rho )}^{(N^{+}j^{-}N^{-})}\mid
A^{(N^{+}j^{-}N^{-})}\parallel E_{(\beta ...\delta \sigma ...\nu ...\tau
)}^{(N^{+}j^{-}N^{-})}>_{g^{(N^{+}N^{-})}}  \notag \\
&=&g_{\alpha \beta }^{(1+)}...g_{\gamma \delta }^{(N+)}g_{\lambda \sigma
}^{(1-)}...A_{\mu \nu }^{(j-)}...g_{\rho \tau }^{(N-)},  \TCItag{4.34}
\end{eqnarray}%
with%
\begin{equation}
<E_{(\alpha ...\gamma \lambda ...\mu ...\rho )}^{(N^{+}j^{-}N^{-})}\mid
=<e_{(\alpha )}^{(1+)}...e_{(\gamma )}^{(N+)}e_{(\lambda )}^{(1-)}...e_{(\mu
)}^{(j-)}...e_{(\rho )}^{(N-)}\mid .  \tag{4.35}
\end{equation}%
Thus, invoking (2.28) and (2.37), we also obtain the values%
\begin{equation}
<\Psi ^{(N^{+}N^{-})}\mid A^{(j^{+}N^{+}N^{-})}\parallel \Psi
^{(N^{+}N^{-})}>_{g^{(N^{+}N^{-})}}=(-1)^{N^{-}}<A^{(j+)}>_{\Psi ^{(j+)}}, 
\tag{4.36a}
\end{equation}%
and%
\begin{equation}
<\Psi ^{(N^{+}N^{-})}\mid A^{(N^{+}j^{-}N^{-})}\parallel \Psi
^{(N^{+}N^{-})}>_{g^{(N^{+}N^{-})}}=(-1)^{n^{-}}<A^{(j-)}>_{\Psi ^{(j-)}}, 
\tag{4.36b}
\end{equation}%
with $n^{-}=N^{-}-1$.

The adjoint densities related to the state (4.20) are set as%
\begin{equation}
\rho ^{(\Psi N^{+}N^{-})}=\rho ^{\lbrack \Psi N^{+}]}\otimes \rho ^{\lbrack
\Psi N^{-}]},\text{ }\rho _{(\Psi N^{+}N^{-})}^{\ast }=\rho _{\lbrack \Psi
N^{+}]}^{\ast }\otimes \rho _{\lbrack \Psi N^{-}]}^{\ast },  \tag{4.37}
\end{equation}%
with the contribution%
\begin{equation}
\rho ^{\lbrack \Psi N^{\pm }]}=\overset{N^{\pm }}{\underset{k=1}{\otimes }}%
g^{(k\pm )}\mid \Psi ^{(k\pm )}><\Psi ^{(k\pm )}\mid ,  \tag{4.38}
\end{equation}%
which bears pseudo Hermiticity and satisfies relations of the form of (4.6)
and (4.7). By recalling (4.2), we deduce the eigenvalue equation%
\begin{equation}
<\Psi ^{\lbrack N^{\pm }]}\mid \rho ^{\lbrack \Psi N^{\pm }]}=(+1)<\Psi
^{\lbrack N^{\pm }]}\mid ,  \tag{4.39}
\end{equation}%
which produces the value%
\begin{equation}
<\Psi ^{\lbrack N^{\pm }]}\mid \rho ^{\lbrack \Psi N^{\pm }]}\parallel \Psi
^{\lbrack N^{\pm }]}>_{g^{[N^{\pm }]}}=(\pm 1)^{N^{\pm }}.  \tag{4.40}
\end{equation}

The canonical representation of $\rho ^{\lbrack \Psi N^{\pm }]}$ is supplied
by the computational result%
\begin{eqnarray}
&<&e_{(\mu )}^{(1\pm )}...e_{(\nu )}^{(N\pm )}\mid \rho ^{\lbrack \Psi
N^{\pm }]}\parallel e_{(\lambda )}^{(1\pm )}...e_{(\sigma )}^{(N\pm
)}>_{g^{[N^{\pm }]}}  \notag \\
&=&\Delta _{\mu \rho }^{(1\pm )}\overline{\Psi _{(1\pm )}^{\rho }}...\Delta
_{\nu \tau }^{(N\pm )}\overline{\Psi _{(N\pm )}^{\tau }}\Psi _{\lambda
}^{(1\pm )}...\Psi _{\sigma }^{(N\pm )},  \TCItag{4.41}
\end{eqnarray}%
while the representation of $\rho ^{(\Psi N^{+}N^{-})}$ comes from%
\begin{eqnarray}
&<&E_{(\alpha ...\mu \lambda ...\rho )}^{(N^{+}N^{-})}\mid \rho ^{(\Psi
N^{+}N^{-})}\parallel E_{(\beta ...\nu \sigma ...\tau
)}^{(N^{+}N^{-})}>_{g^{(N^{+}N^{-})}}  \notag \\
&=&<e_{(\alpha )}^{(1+)}...e_{(\mu )}^{(N+)}\mid \rho ^{\lbrack \Psi
N^{+}]}\parallel e_{(\beta )}^{(1+)}...e_{(\nu )}^{(N+)}>_{g^{[N^{+}]}} 
\notag \\
\times &<&e_{(\lambda )}^{(1-)}...e_{(\rho )}^{(N-)}\mid \rho ^{\lbrack \Psi
N^{-}]}\parallel e_{(\sigma )}^{(1-)}...e_{(\tau )}^{(N-)}>_{g^{[N^{-}]}}, 
\TCItag{4.42}
\end{eqnarray}%
where the $\Delta $-entries carry the particle-antiparticle labels of
(4.31), and the $E^{(N^{+}N^{-})}$-basis may be obtained from (4.32) by
hiding the $(j^{+})$-factor. It follows from (4.40) that%
\begin{equation}
<\Psi ^{(N^{+}N^{-})}\mid \rho ^{(\Psi N^{+}N^{-})}\parallel \Psi
^{(N^{+}N^{-})}>_{_{g^{(N^{+}N^{-})}}}=(-1)^{N^{-}}.  \tag{4.43}
\end{equation}

Conditional, mutual and relative entropies for the composite densities we
have constructed can all be formally defined in the same way as in the
non-relativistic context. The same property applies as well to tensor
products of densities like that given as (4.15).

We can evaluate the traces of the densities (4.37) by simply allowing for a
trivial version of the rule (2.44). We have, in effect,%
\begin{equation}
\text{Tr }\rho ^{\lbrack \Psi N^{\pm }]}=<\Psi ^{\lbrack N^{\pm }]}\mid
g^{[N^{\pm }]}\parallel \Psi ^{\lbrack N^{\pm }]}>_{g^{[N^{\pm }]}}=+1, 
\tag{4.44}
\end{equation}%
and%
\begin{equation}
\text{Tr }\rho ^{(\Psi N^{+}N^{-})}=\text{Tr }\rho ^{\lbrack \Psi N^{+}]}%
\text{Tr }\rho ^{\lbrack \Psi N^{-}]},  \tag{4.45}
\end{equation}%
with Tr $\rho ^{\lbrack \Psi N^{\pm }]}=$Tr $\rho _{\lbrack \Psi N^{\pm
}]}^{\ast }$. The traces that occur on the right-hand side of (4.45) can be
calculated explicitly by combining (4.41) and the prescription%
\begin{equation}
\text{Tr }\rho ^{\lbrack \Psi N^{\pm }]}=<E_{(\mu ...\nu )}^{(N^{\pm })}\mid
\rho ^{\lbrack \Psi N^{\pm }]}\parallel E_{(\lambda ...\sigma )}^{(N^{\pm
})}>_{g^{[N^{\pm }]}}g_{(1\pm )}^{\ast \lambda \mu }...g_{(N\pm )}^{\ast
\sigma \nu },  \tag{4.46}
\end{equation}%
where%
\begin{equation}
<E_{(\mu ...\nu )}^{(N^{\pm })}\mid \doteqdot <e_{(\mu )}^{(1\pm
)}...e_{(\nu )}^{(N\pm )}\mid .  \tag{4.47}
\end{equation}%
We thus obtain the formal expression%
\begin{equation}
\text{Tr }\rho ^{\lbrack \Psi N^{\pm }]}=\overset{N^{\pm }}{\underset{k=1}{%
\Pi }}<<\Psi ^{(k\pm )}\mid \Psi ^{(k\pm )}>>=1.  \tag{4.48}
\end{equation}

For any of the states $<\Psi ^{(\eta \pm )}\mid $ carried by (4.20), a
dynamical reduction of (4.37) may be defined by taking the partial traces
over the $(\eta \pm )$-subsystems. Loosely speaking, tracing out some
subsystem drops the respective states from the former density
configurations. Formally, we have%
\begin{equation}
\text{Tr}_{(\eta \pm )}\text{ }\rho ^{\lbrack \Psi N^{\pm }]}\doteqdot <\Psi
^{(\eta \pm )}\mid g^{(\eta \pm )}\parallel \Psi ^{(\eta \pm )}>_{g^{(\eta
\pm )}}\rho ^{(\eta ^{\pm }\Psi N^{\pm })},  \tag{4.49}
\end{equation}%
with $\rho ^{(\eta ^{\pm }\Psi N^{\pm })}$ accordingly denoting here for
once the densities that are constituted by the remaining states of $\rho
^{\lbrack \Psi N^{\pm }]}$. Then, tracing out all the subsystems described
by (4.38), except the $(j\pm )$-ones, takes us back to the prescription
(4.8) through%
\begin{equation}
<A^{(j\pm )}>_{\Psi ^{(j\pm )}}=\text{Tr }(g^{(j\pm )}\rho ^{(\Psi j\pm
)}A^{(j\pm )}).  \tag{4.50}
\end{equation}

\section{Local projective measurements}

In this Section, we will construct the structures associated to the
projective measurements which should be tied in with the framework of
Sections 3 and 4. Our constructions amount to a $\mathcal{P}_{+}^{\uparrow }$%
-covariant version of the projective ones borne by the old quantum
mechanical formulation [19]. We postulate that the reduction and destruction
of single and composite states caused by ordinary projective measurements in
the absence of spectral degeneracy, may be made into invariant features of
projective observations. For convenience, we will construct the
corresponding configurations in the frame that takes up the canonical bases.
The adaptation to our context of the measurement algebra developed by
Schwinger [20] will hopefully be carried out separately in another paper.

We allow for the state for the pair $(p^{+},p^{-})$ as given by (3.2). Any
projective measurements on $p^{\pm }$ are characterized by pseudo-Hermitian
operator restrictions that obey defining prescriptions like%
\begin{equation}
<\Phi ^{(p\pm )}\mid \pi _{(\mu )}^{(p\pm )}=<\Phi _{(\mu )}^{(p\pm )}\mid ,%
\text{ }<\Phi _{(p\pm )}^{\ast }\mid \pi _{(p\pm )}^{\ast (\mu )}=<\Phi
_{(p\pm )}^{\ast (\mu )}\mid ,  \tag{5.1a}
\end{equation}%
with%
\begin{equation}
<\Phi _{(0)}^{(p\pm )}\mid =\Phi _{(p\pm )}^{0}<e_{(0)}^{(p\pm )}\mid ,\text{
}<\Phi _{(1)}^{(p\pm )}\mid =\Phi _{(p\pm )}^{1}<e_{(1)}^{(p\pm )}\mid . 
\tag{5.1b}
\end{equation}%
The normalizability of states as prescribed in Section 3 is thus lost when
projective measurements are actually performed. It follows that, calling for
the linear combination%
\begin{equation}
<\Phi ^{(p\pm )}\mid \pi _{(\mu )}^{(p\pm )}=\Phi _{(p\pm )}^{\lambda }\pi
_{(\mu )\lambda }^{(p\pm )}{}^{\sigma }<e_{(\sigma )}^{(p\pm )}\mid , 
\tag{5.2}
\end{equation}%
together with its adjoint version, yields the representations%
\begin{equation}
(\pi _{(0)\lambda \sigma }^{(p\pm )}{})=%
\begin{pmatrix}
\pm 1 & 0 \\ 
0 & 0%
\end{pmatrix}%
=(\pi _{(p\pm )}^{\ast (0)\lambda \sigma }{}),  \tag{5.3a}
\end{equation}%
and%
\begin{equation}
(\pi _{(1)\lambda \sigma }^{(p\pm )}{})=%
\begin{pmatrix}
0 & 0 \\ 
0 & \pm 1%
\end{pmatrix}%
=(\pi _{(p\pm )}^{\ast (1)\lambda \sigma }{}),  \tag{5.3b}
\end{equation}%
whence Tr $\pi _{(\mu )}^{(p\pm )}{}=$Tr $\pi _{(p\pm )}^{\ast (\mu )}{}=+1$%
. A straightforward calculation then gives the values%
\begin{equation}
<\Phi ^{(p\pm )}\mid \pi _{(\mu )}^{(p\pm )}\parallel \Phi ^{(p\pm
)}>_{g^{\pm }}=\pm w_{(\mu )}^{(p\pm )},  \tag{5.4}
\end{equation}%
along with the ones for $\pi _{(p\pm )}^{\ast (\mu )}{}$ (see (3.7)). We
have the property%
\begin{equation}
\pi _{(\mu )}^{(p\pm )}\pi _{(\nu )}^{(p\pm )}=\Delta _{\mu \nu }^{\pm }\pi
_{(\nu )}^{(p\pm )}\text{ (no summation over here)},  \tag{5.5}
\end{equation}%
such that%
\begin{equation}
<\Phi _{(\mu )}^{(p\pm )}\mid \Phi _{(\nu )}^{(p\pm )}>_{g^{\pm }}=g_{\mu
\nu }^{\pm }w_{(\nu )}^{(p\pm )}\text{ (no summation over here)}.  \tag{5.6}
\end{equation}%
Therefore, the possible unstarred reduced states produced by the $\pi $%
-measurement processes appear as the normalized patterns%
\begin{equation}
\frac{1}{\sqrt{w_{(0)}^{(p\pm )}}}<\Phi _{(0)}^{(p\pm )}\mid ,\text{ }\frac{1%
}{\sqrt{w_{(1)}^{(p\pm )}}}<\Phi _{(1)}^{(p\pm )}\mid .  \tag{5.7}
\end{equation}

Equation (2.40) and its adjoint afford a natural form of canonical
decompositions for identity operators. Such configurations may also supply
invariant decompositions for projective-measurement operators, whence the
definition (4.1) gives rise to the completeness property%
\begin{equation}
\text{Tr }(\rho ^{(\Phi \pm )}\mid e_{(\mu )}^{(p\pm )}>g_{\pm }^{\ast \mu
\nu }<e_{(\nu )}^{(p\pm )}\mid )=w_{(0)}^{(p\pm )}+w_{(1)}^{(p\pm )}. 
\tag{5.8}
\end{equation}%
From (5.7), we see that successive projective measurements on constituents
of composite systems can be performed by implementing selection procedures
similar to that we had utilized in the preceding Section for introducing
(4.30) and (4.33). The description of the measurement processes for any
observables thus gets completed when the relevant spectral decompositions
are coupled to the prepared states to be dealt with. By taking a
decomposition for $A^{(j\pm )}$, for instance, like the one carried by
(3.26), and supposing that it does not bear degeneracy, we then recover
(4.50) as an expectation prescription of the form of (3.28), namely,%
\begin{equation}
<\Psi ^{(j\pm )}\mid A^{(j\pm )}\parallel \Psi ^{(j\pm )}>_{g^{(j\pm )}}=\pm
a_{(j\pm )}^{\mu }w_{(\mu )}^{(j\pm )}.  \tag{5.9}
\end{equation}%
Hence, after the $\pi _{(\mu )}^{(j\pm )}$-measurements are performed upon
the $(j\pm )$-subsystems, we may use a notation of the type of (4.25) to get
the state reductions%
\begin{equation}
<\Psi ^{\lbrack N^{\pm }]}\mid \pi _{(0)}^{(j\pm )}=<\Psi ^{(1\pm )}...\frac{%
1}{\sqrt{w_{(0)}^{(j\pm )}}}\Psi _{(0)}^{(j\pm )}...\Psi ^{(N\pm )}\mid , 
\tag{5.10a}
\end{equation}%
and\footnote{%
No state reductions happen when projective measurements of degenerate
spectra are carried out.}%
\begin{equation}
<\Psi ^{\lbrack N^{\pm }]}\mid \pi _{(1)}^{(j\pm )}=<\Psi ^{(1\pm )}...\frac{%
1}{\sqrt{w_{(1)}^{(j\pm )}}}\Psi _{(1)}^{(j\pm )}...\Psi ^{(N\pm )}\mid . 
\tag{5.10b}
\end{equation}

\section{The Poincar\'{e} subgroup of $SU(2,2)$}

From an algebraic viewpoint, $SU(2,2)$ is the special (unimodular) group
constituted by the usual operation of matrix multiplication and the set of
matrices that represent either of the totalities $\{\mathcal{U},\mathcal{U}%
^{\ast }\}$ of pseudo-unitary operators in $\mathfrak{C}$ and $\mathfrak{C}%
^{\ast }$. It should be stressed that the only admissible basis devices for
representing pseudo-unitary operators in $\mathfrak{C}$ and $\mathfrak{C}%
^{\ast }$ are $SU(2,2)$-related to each other. In this Section, we will thus
drop the $(p\pm )$-labels from inner products.

If $u\in \mathcal{U}$, we may then write the configuration%
\begin{equation}
uu^{\bigstar }=I\Leftrightarrow ugu^{\dag }=g,  \tag{6.1}
\end{equation}%
along with its adjoint version. Hence, utilizing a decomposition for each of 
$u$ and $u^{\bigstar }$ of the same type as that given by (2.15), yields the 
$\bigstar $-invariant operator relations%
\begin{equation}
u^{++}u^{++\bigstar }+u^{+-}u^{+-\bigstar }=I^{+},\text{ }%
u^{--}u^{--\bigstar }+u^{-+}u^{-+\bigstar }=I^{-},  \tag{6.2a}
\end{equation}%
and%
\begin{equation}
u^{++}u^{-+\bigstar }+u^{+-}u^{--\bigstar }=0=u^{-+}u^{++\bigstar
}+u^{--}u^{+-\bigstar },  \tag{6.2b}
\end{equation}%
where $I^{\pm }\doteqdot $ Res$I_{\mathfrak{C}^{\pm }}$. For the
representation of the relations (6.2), we have the defining entry constraints%
\begin{equation}
u_{\mu \lambda }^{++}g_{+}^{\ast \lambda \sigma }{}u_{\sigma \nu
}^{++\bigstar }+u_{\mu \lambda }^{+-}g_{-}^{\ast \lambda \sigma }{}u_{\sigma
\nu }^{+-\bigstar }=g_{\mu \nu }^{+},  \tag{6.3a}
\end{equation}%
and%
\begin{equation}
u_{\mu \lambda }^{--}g_{-}^{\ast \lambda \sigma }{}u_{\sigma \nu
}^{--\bigstar }+u_{\mu \lambda }^{-+}g_{+}^{\ast \lambda \sigma }{}u_{\sigma
\nu }^{-+\bigstar }=g_{\mu \nu }^{-},  \tag{6.3b}
\end{equation}%
along with%
\begin{equation}
u_{\mu \lambda }^{++}g_{+}^{\ast \lambda \sigma }{}u_{\sigma \nu
}^{-+\bigstar }+u_{\mu \lambda }^{+-}g_{-}^{\ast \lambda \sigma }{}u_{\sigma
\nu }^{--\bigstar }=0_{2},  \tag{6.3c}
\end{equation}%
and the $\bigstar $-conjugate of (6.3c). The matrix entries for $u$ in any
admissible basis are expressed in much the same way as those of (2.17),
whilst the ones for $u^{\bigstar }$ can be obtained by invoking the
interchanges of operator actions that had been used for setting up (3.44).
For instance,%
\begin{equation}
u_{\mu \nu }^{+-\bigstar }=<e_{(\mu )}^{-}\mid u^{+-\bigstar }\parallel
e_{(\nu )}^{+}>_{g^{+}}=g_{\mu \lambda }^{+}\Delta _{+}^{\ast \lambda \sigma
}u_{\sigma \rho }^{+-\dag }g_{+}^{\ast \rho \tau }{}\Delta _{\tau \nu }^{+},
\tag{6.4}
\end{equation}%
and%
\begin{equation}
u_{\mu \nu }^{-+\bigstar }=<e_{(\mu )}^{+}\mid u^{-+\bigstar }\parallel
e_{(\nu )}^{-}>_{g^{-}}=g_{\mu \lambda }^{-}\Delta _{-}^{\ast \lambda \sigma
}u_{\sigma \rho }^{-+\dag }g_{-}^{\ast \rho \tau }{}\Delta _{\tau \nu }^{-}.
\tag{6.5}
\end{equation}

The $(2\times 2)$-blocks $A$, $a$, $b$ and $B$ of Ref. [16] that correspond
to (6.3) may be related to the matrix contributions formed by $u_{\mu
\lambda }^{++\dag }$, $u_{\mu \lambda }^{+-\dag }$, $u_{\mu \lambda
}^{-+\dag }$ and $u_{\mu \lambda }^{--\dag }$, respectively. Whenever $u$ is
taken to bear unitarity as well, its decomposition turns out to be such that
the constituents $u^{+-}$ and $u^{-+}$ amount both to zero operators. In
this case, we should thus take account of the conditions%
\begin{equation}
u^{++\bigstar }=u^{++\dag },\text{ }u^{--\bigstar }=u^{--\dag },  \tag{6.6a}
\end{equation}%
and [30]%
\begin{equation}
(u_{\mu \nu }^{++})\in U(2)\ni (u_{\mu \nu }^{--}),\text{ }\det (u_{\mu \nu
}^{++})=\exp [i\phi ]=\det (u_{\mu \nu }^{--})^{-1},  \tag{6.6b}
\end{equation}%
with $\phi $ being some real number.

Equation (6.1) constitutes what is called the $g$-realization of $SU(2,2)$.
Another greatly interesting realization of this group [16] takes up the Gram
operator specified as%
\begin{equation}
G:(\Lambda _{+}^{0},\Lambda _{+}^{1},\Lambda _{-}^{0},\Lambda
_{-}^{1})\mapsto (\Lambda _{-}^{0},\Lambda _{-}^{1},\Lambda _{+}^{0},\Lambda
_{+}^{1}),  \tag{6.7}
\end{equation}%
which is invariantly represented by%
\begin{equation}
(G_{\mu \nu })=\left( 
\begin{array}{ll}
0_{2} & I_{2} \\ 
I_{2} & 0_{2}%
\end{array}%
\right) =(G^{\ast \mu \nu }).  \tag{6.8}
\end{equation}%
For the defining constraints for the $G$-realization, we have the
prescription%
\begin{equation}
UU^{[\bigstar ]}=I\Leftrightarrow UGU^{\dag }=G,  \tag{6.9a}
\end{equation}%
whose representation satisfies\footnote{%
In (6.9b), we must take $U_{\mu \nu }^{\dag }=<e_{(\mu )}\mid U^{\dag
}\parallel e_{(\nu )}>_{G}$.}%
\begin{equation}
U_{\mu \lambda }G^{\ast \lambda \sigma }U_{\sigma \nu }^{[\bigstar ]}=G_{\mu
\nu }\Leftrightarrow U_{\mu \lambda }\Delta ^{\ast \lambda \sigma }U_{\sigma
\nu }^{\dag }=\Delta _{\mu \nu },  \tag{6.9b}
\end{equation}%
where the $\bigstar $-symbol in square brackets denotes the pseudo-Hermitian
conjugation with respect to the $G$-inner product, and%
\begin{equation}
(U_{\mu \nu }{})=\left( 
\begin{array}{ll}
a & A \\ 
B & b%
\end{array}%
\right) ,\text{ }U_{\mu \nu }\doteqdot U_{\mu }{}^{\lambda }G_{\lambda \nu }.
\tag{6.10}
\end{equation}%
As displayed in Ref. [16], the $G$-realization $(2\times 2)$-blocks $A$, $a$%
, $b$ and $B$ make up the matrix $(U_{\mu }^{\dag }{}^{\nu })$. When
considered adequately, the products of the blocks carried by $(U_{\mu \nu
}{})$ fulfill a skew-Hermiticity property. If $U$ bears unitarity too, we
have to account for block matrices prescribed as%
\begin{equation}
(U_{\mu \nu }{})=\left( 
\begin{array}{ll}
a & A \\ 
A & a%
\end{array}%
\right) ,\text{ }AA^{\dag }+aa^{\dag }=I_{2},\text{ }Aa^{\dag }=-aA^{\dagger
},  \tag{6.11}
\end{equation}%
in which case $U^{[\bigstar ]}=U^{\dag }$.

The primary relationship involving the $gG$-realizations is afforded by%
\begin{equation}
\mathfrak{M}^{-1}G\mathfrak{M}=g,  \tag{6.12a}
\end{equation}%
which supplies the operator statement%
\begin{equation}
\mathfrak{M}^{-1}U\mathfrak{M}=u\Rightarrow \mathfrak{M}^{-1}U^{[\bigstar ]}%
\mathfrak{M}=u^{\bigstar },  \tag{6.12b}
\end{equation}%
where $\mathfrak{M}$ stands for a unitary operator that does not admit any
representative matrix from either realization. As a consequence of (6.12),
we have the determinant-preserving correlation%
\begin{equation}
U_{\mu \nu }=\mathfrak{M}_{\mu }{}^{\lambda }u_{\lambda \sigma }\overline{%
\mathfrak{M}{}^{\sigma }{}_{\nu }}.  \tag{6.13}
\end{equation}%
A particularly useful matrix for $\mathfrak{M}$ appears as%
\begin{equation}
(\mathfrak{M}_{\mu }{}^{\nu })=\frac{1}{\sqrt{2}}\left( 
\begin{array}{ll}
I_{2} & -I_{2} \\ 
I_{2} & I_{2}%
\end{array}%
\right) .  \tag{6.14}
\end{equation}

In the $G$-realization, the representation of the group $\mathcal{P}%
_{+}^{\uparrow }$ consists of all ten-parameter $SU(2,2)$-matrices of the
form [31]%
\begin{equation}
(U_{\mu \nu }^{(\mathcal{P}_{+}^{\uparrow })}{})=\left( 
\begin{array}{ll}
iWa^{-1\dag } & a \\ 
a^{-1\dag } & 0_{2}%
\end{array}%
\right) ,  \tag{6.15a}
\end{equation}%
where $a$ belongs to $SL(2,\mathbf{C})$ and essentially represents an
element of $\mathcal{L}_{+}^{\uparrow }$, while $W$ is the van der Waerden
[10] Hermitian $(2\times 2)$-matrix associated to a time-like or space-like
Minkowskian translation (see (6.21) below). Any null Minkowskian translation
yields $\det W=0$, and has been ruled out by this point. The entries of the
Poincar\'{e} matrices exhibited in Refs. [31] may be taken to equal the ones
of 
\begin{equation}
U_{\mu }^{(\mathcal{P}_{+}^{\uparrow })}{}^{\nu }=U_{\mu \lambda }^{(%
\mathcal{P}_{+}^{\uparrow })}{}G^{\ast \lambda \nu }.  \tag{6.15b}
\end{equation}%
We observe that the explicit $i$-factor carried by the right-hand side of
(6.15a) just ensures the required skew Hermiticity of the product $%
ia^{-1}Wa^{-1\dag }$. By employing (6.13) and (6.14), we write the $g$%
-realization version of the matrix (6.15a) as%
\begin{equation}
(u_{\mu \nu }^{(\mathcal{P}_{+}^{\uparrow })}{})=\frac{1}{2}\left( 
\begin{array}{cc}
a+(I_{2}+iW)a^{-1\dagger } & a-(I_{2}+iW)a^{-1\dagger } \\ 
-a+(I_{2}-iW)a^{-1\dagger } & -a-(I_{2}-iW)a^{-1\dagger }%
\end{array}%
\right) .  \tag{6.16}
\end{equation}%
The representation of $\mathcal{L}_{+}^{\uparrow }$ is thereupon formed by
the set of six-parameter configurations of the type%
\begin{equation}
(U_{\mu \nu }^{(\mathcal{L}_{+}^{\uparrow })}{})=\left( 
\begin{array}{ll}
0_{2} & a \\ 
a^{-1\dag } & 0_{2}%
\end{array}%
\right) ,  \tag{6.17}
\end{equation}%
and%
\begin{equation}
(u_{\mu \nu }^{(\mathcal{L}_{+}^{\uparrow })}{})=\frac{1}{2}\left( 
\begin{array}{cc}
a+a^{-1\dagger } & a-a^{-1\dagger } \\ 
-a+a^{-1\dagger } & -a-a^{-1\dagger }%
\end{array}%
\right) .  \tag{6.18}
\end{equation}

When the unitary intersection (3.12) is called for, we must replace (6.16)
with%
\begin{equation}
(u_{\mu \nu }^{(\mathcal{P}_{+}^{\uparrow })}{})=\frac{1}{\sqrt{2}}\left( 
\begin{array}{cc}
(I_{2}+iW)\beta & 0_{2} \\ 
0_{2} & \beta ^{\dag }(I_{2}-iW)%
\end{array}%
\right) ,  \tag{6.19}
\end{equation}%
where $W$ has now to be normalized as%
\begin{equation}
W^{2}=I_{2},\text{ }\det W=\pm 1,  \tag{6.20}
\end{equation}%
and $\beta \in U(2)$. It is clear that the matrix (6.19) agrees with the
conditions (6.6). Therefore, if a Hermitian matrix corresponds to some
non-null spacetime translation and enters a unitary Poincar\'{e} element of
the $g$-realization of $SU(2,2)$, then it may be associated with a
normalized world vector like%
\begin{equation}
\tau ^{a}=\frac{1}{\sqrt{2}}T^{a},\text{ }T^{b}T_{b}=2\det W.  \tag{6.21a}
\end{equation}%
Lower-case Latin indices have been used here for labelling the components of
spacetime translations like, for instance, $T^{0}$, $T^{1}$, $T^{2}$ and $%
T^{3}$. Some calculations thus produce the formulae%
\begin{equation}
\det W=+1\Rightarrow \det [\frac{1}{\sqrt{2}}(I_{2}\pm iW)]=\pm \frac{i}{%
\sqrt{2}}T^{0},  \tag{6.21b}
\end{equation}%
with $T^{0}=\pm \sqrt{2}$, and\footnote{%
Due to the relations (6.21), the only admissible translations $\tau ^{a}$
are of the types $(\pm 1,0,0,0)$ and $(0,\tau ^{1},\tau ^{2},\tau ^{3})$.
This property is passed on to any realization of $SU(2,2)$.}%
\begin{equation}
\det W=-1\Rightarrow \det [\frac{1}{\sqrt{2}}(I_{2}\pm iW)]=1\pm \frac{i}{%
\sqrt{2}}T^{0},  \tag{6.21c}
\end{equation}%
with $T^{0}=0$. Hence, the matrix (6.19) carries seven real parameters
whereas its $\mathcal{L}_{+}^{\uparrow }$-version emerges as the
four-parameter structure%
\begin{equation}
(u_{\mu \nu }^{(\mathcal{L}_{+}^{\uparrow })}{})=\left( 
\begin{array}{cc}
\beta & 0_{2} \\ 
0_{2} & \beta ^{\dag }%
\end{array}%
\right) .  \tag{6.22}
\end{equation}%
The $G$-version of (6.19) is accordingly given by the pattern (6.11)
together with the identifications%
\begin{equation}
a=\frac{1}{2\sqrt{2}}[(I_{2}+iW)\beta +\beta ^{\dag }(I_{2}-iW)], 
\tag{6.23a}
\end{equation}%
and%
\begin{equation}
A=\frac{1}{2\sqrt{2}}[(I_{2}+iW)\beta -\beta ^{\dag }(I_{2}-iW)], 
\tag{6.23b}
\end{equation}%
whilst (6.22) similarly yields%
\begin{equation}
(U_{\mu \nu }^{(\mathcal{L}_{+}^{\uparrow })}{})=\frac{1}{2}\left( 
\begin{array}{cc}
\beta +\beta ^{\dag } & \beta -\beta ^{\dag } \\ 
\beta -\beta ^{\dag } & \beta +\beta ^{\dag }%
\end{array}%
\right) .  \tag{6.24}
\end{equation}

\section{Observational correlations}

Any changes of spacetime frames are induced by the action of the dynamical
subgroup $\mathcal{P}_{Dyn}^{+\uparrow }$ of $SU(2,2)\cap U(4)$ which
consists of the totality of Poincar\'{e} matrices of the form (6.19) whose $%
\beta $-pieces represent either boosts along arbitrary spacetime directions
or proper rotations.\footnote{%
In such cases, the matrix (6.22) turns out to carry three real parameters.}
By virtue of (6.6), the relationships between any copies of the
computational bases for different frames, and also the behaviours of any
physical entities, are effectively controlled by pseudo-unitary restrictions
that enter into arrays like%
\begin{equation}
u_{Dyn}{}=\left( 
\begin{array}{ll}
u_{Dyn}^{+}{} & 0 \\ 
0 & u_{Dyn}^{-}{}%
\end{array}%
\right) ,\text{ }u_{Dyn}^{\pm \bigstar }{}=(u_{Dyn}^{\pm
}{})^{-1}=u_{Dyn}^{\pm \dag }{}.  \tag{7.1}
\end{equation}%
These restrictions are then represented by six-real-parameter matrices $%
\{(u_{\mu \nu }^{Dyn}{})\}$ subject to%
\begin{equation}
\mathcal{P}_{Dyn}^{+\uparrow }\ni (u_{\mu \nu }^{Dyn}{}),\text{ }\beta \in
SL(2,\mathbf{C})\cap U(2),  \tag{7.2}
\end{equation}%
with the entry prototype%
\begin{equation}
u_{\mu \nu }^{Dyn\pm }{}\doteqdot u_{\mu \nu }^{\pm }{}=<e_{(\mu )}^{(p\pm
)}\mid u_{Dyn}^{\pm }{}\parallel e_{(\nu )}^{(p\pm )}>_{g^{\pm }}.  \tag{7.3}
\end{equation}

Equation (7.1) yields the invariance under $\mathcal{P}_{Dyn}^{+\uparrow }$
of both $g_{\mu \nu }^{\pm }$ and $\Delta _{\mu \nu }^{\pm }$. In effect, we
have%
\begin{equation}
g_{\mu \nu }^{\prime \pm }=<e_{(\mu )}^{(p\pm )}\mid u_{Dyn}^{\pm }{}\mid
u_{Dyn}^{\pm }{}\mid e_{(\nu )}^{(p\pm )}>_{g^{\pm }}=g_{\mu \nu }^{\pm }, 
\tag{7.4}
\end{equation}%
and%
\begin{equation}
\Delta _{\mu \nu }^{\prime \pm }=<<e_{(\mu )}^{(p\pm )}\mid u_{Dyn}^{\pm
}{}\mid u_{Dyn}^{\pm }{}\mid e_{(\nu )}^{(p\pm )}>>=\Delta _{\mu \nu }^{\pm
},  \tag{7.5}
\end{equation}%
with the primed kernel letters thus referring to the frame of $\mathcal{L}%
_{+}^{\uparrow }$ which carries the computational basis%
\begin{equation}
<e_{(\mu )}^{\prime (p\pm )}\mid =<e_{(\mu )}^{(p\pm )}\mid u_{Dyn}^{\pm }{}.
\tag{7.6}
\end{equation}%
It can therefore be said that the decompositions which involve the spaces of
state vectors for any particle-antiparticle pairs, provide invariant
prescriptions. It is evident that (7.4) and (7.5) may be reset as%
\begin{equation}
g_{\mu \nu }^{\prime \pm }=u_{\mu \lambda }^{\pm }g_{\pm }^{\ast \lambda
\sigma }{}u_{\sigma \nu }^{\pm \bigstar }=g_{\mu \nu }^{\pm },  \tag{7.7a}
\end{equation}%
and%
\begin{equation}
\Delta _{\mu \nu }^{\prime \pm }=u_{\mu \lambda }^{\pm }\Delta _{\pm }^{\ast
\lambda \sigma }{}u_{\sigma \nu }^{\pm \dag }=\Delta _{\mu \nu }^{\pm }. 
\tag{7.7b}
\end{equation}%
The behaviours of the adjoint versions of $g_{\mu \nu }^{\pm }$ and $\Delta
_{\mu \nu }^{\pm }$ have to be specified by%
\begin{equation}
g_{\pm }^{\ast \prime \mu \nu }{}=u_{\pm }^{\ast \mu \lambda }g_{\lambda
\sigma }^{\pm }u_{\pm }^{\ast \bigstar \sigma \nu }=g_{\pm }^{\ast \mu \nu
}{},  \tag{7.8a}
\end{equation}%
and%
\begin{equation}
\Delta _{\pm }^{\ast \prime \mu \nu }{}=u_{\pm }^{\ast \mu \lambda }\Delta
_{\lambda \sigma }^{\pm }u_{\pm }^{\ast \dag \sigma \nu }=\Delta _{\pm
}^{\ast \mu \nu }{},  \tag{7.8b}
\end{equation}%
with\footnote{%
From (7.9), we also get $u_{\pm }^{\ast \mu \lambda }u_{\lambda \nu }^{\pm
\bigstar }=\Delta _{\pm }^{\ast \mu \lambda }\Delta _{\lambda \nu }^{\pm
}=u_{\pm }^{\ast \bigstar \mu \lambda }u_{\lambda \nu }^{\pm }$.}%
\begin{equation}
u_{\mu \lambda }^{\pm }u_{\pm }^{\ast \bigstar \lambda \nu }=\Delta _{\mu
\lambda }^{\pm }\Delta _{\pm }^{\ast \lambda \nu }=u_{\mu \lambda }^{\pm
\bigstar }u_{\pm }^{\ast \lambda \nu }.  \tag{7.9}
\end{equation}

In fact, the invariance of $g_{\mu \nu }^{\pm }$ and $g_{\pm }^{\ast \mu \nu
}$ as well as the defining group closedness of $SU(2,2)$ with respect to the
ordinary operation of matrix multiplication, permit us to use relations like 
$u_{\mu \nu }^{\pm }=u_{\mu }^{\pm }{}^{\lambda }g_{\lambda \nu }^{\pm }$
and $u_{\mu }^{\pm }{}^{\nu }=u_{\mu \lambda }^{\pm }g_{\pm }^{\ast \lambda
\nu }$ without having to take any choices of frames into consideration. If
single particles or antiparticles are to be considered explicitly, we must
appropriately allow for the form of one of the configurations%
\begin{equation}
\frac{1}{\sqrt{2}}\left( 
\begin{array}{cc}
(I_{2}+iW)\beta & 0_{2} \\ 
0_{2} & 0_{2}%
\end{array}%
\right) ,\text{ }\frac{1}{\sqrt{2}}\left( 
\begin{array}{cc}
0_{2} & 0_{2} \\ 
0_{2} & \beta ^{\dag }(I_{2}-iW)%
\end{array}%
\right) ,  \tag{7.10}
\end{equation}%
which clearly preserve the individual particle-antiparticle characters of
the dynamical descriptions.

Any states like those of (3.2) and (4.20) are naturally invariant under $%
\mathcal{P}_{Dyn}^{+\uparrow }$. This assertion rests upon the fact that the
effective dynamical-group action requires that%
\begin{equation}
<\Phi ^{\prime (p\pm )}\mid \Phi ^{\prime (p\pm )}>_{g^{\prime \pm }}=<\Phi
^{(p\pm )}\mid \Phi ^{(p\pm )}>_{g^{\pm }},  \tag{7.11}
\end{equation}%
and, consequently, we have to demand that%
\begin{equation}
<\Phi _{(p\pm )}^{\prime }\mid =\Phi _{(p\pm )}^{\lambda }u_{\lambda }^{\pm
\bigstar }{}^{\sigma }u_{\sigma }^{\pm }{}^{\tau }<e_{(\tau )}^{(p\pm )}\mid
=<\Phi _{(p\pm )}\mid .  \tag{7.12}
\end{equation}%
It follows that the normalization condition (3.3) is invariant even though
each of the probabilities (3.10) is not. As the specification of the
operator actions of $g^{\pm }$ and $g_{\pm }^{\ast }$ is presumably the same
in any frame, the decomposition (2.40) and the expressions for the density
operators of Section 4, including (4.10), (4.15) and (5.8), bear an
invariant character. For the amplitude (4.22), we have the tensor law%
\begin{equation}
C_{(N^{+}N^{-})}^{\prime \mu ...\nu \lambda ...\sigma
}=C_{(N^{+}N^{-})}^{\xi ...\zeta \rho ...\tau }u_{\xi }^{+\bigstar }{}^{\mu
}...u_{\zeta }^{+\bigstar }{}^{\nu }u_{\rho }^{-\bigstar }{}^{\lambda
}...u_{\tau }^{-\bigstar }{}^{\sigma }.  \tag{7.13}
\end{equation}

The behaviours under $\mathcal{P}_{Dyn}^{+\uparrow }$ of dynamical variables
and spectra depend closely on the physical nature of the magnitudes being
observed. For bringing out the immediately relevant situations, we express
(3.18a) in the primed frame as%
\begin{equation}
A_{\mu \nu }^{\prime (p\pm )}=<e_{(\mu )}^{\prime (p\pm )}\mid A^{\prime
(p\pm )}\parallel e_{(\nu )}^{\prime (p\pm )}>_{g^{\prime \pm }},  \tag{7.14}
\end{equation}%
and write down the correlation%
\begin{equation}
A^{\prime (p\pm )}=A^{(p\pm )}\Rightarrow A_{\mu \nu }^{\prime (p\pm
)}=u_{\mu \lambda }^{\pm }{}A_{(p\pm )}^{\ast \lambda \sigma }u_{\sigma \nu
}^{\pm \bigstar },  \tag{7.15}
\end{equation}%
along with%
\begin{equation}
A_{\mu \nu }^{\prime (p\pm )}=A_{\mu \nu }^{(p\pm )}\Rightarrow A^{\prime
(p\pm )}=u_{Dyn}^{\pm \bigstar }{}A^{(p\pm )}u_{Dyn}^{\pm }{}.  \tag{7.16}
\end{equation}%
Thus, in any situation where both of these correlations hold, we will have
to put into effect the commutativity property%
\begin{equation}
u_{Dyn}^{\pm }{}A^{(p\pm )}=A^{(p\pm )}u_{Dyn}^{\pm }{}.  \tag{7.17}
\end{equation}

Equations (7.15) and (7.16) apply formally to non-observables as well.
Because of (7.12), the projection operators $P^{\pm }$ defined in Section 2
have to bear $\mathcal{P}_{Dyn}^{+\uparrow }$-invariance. The laws
(7.7)-(7.9) produce the invariance of any traces since the adjoint version
of (7.15) amounts to%
\begin{equation}
A_{(p\pm )}^{\ast \prime }=A_{(p\pm )}^{\ast }\Rightarrow A_{(p\pm )}^{\ast
\prime \mu \nu }=u_{\pm }^{\ast \mu \lambda }A_{\lambda \sigma }^{(p\pm
)}u_{\pm }^{\ast \bigstar \sigma \nu }.  \tag{7.18}
\end{equation}%
Hence, (2.41b) establishes the equivalence between the correlations (7.15)
and (7.18). By calling upon (3.25), we also demonstrate that every
degenerate spectrum is invariant. Consequently, the property (7.17) is
deemed to apply to any $A^{(p\pm )}$ whose spectrum bears degeneracy. In
Ref. [17], the $SU(2,2)$-behaviours of operators, matrix elements and traces
were specified apart from any physical consideration.

When the correlation (7.15) holds alone, the value (3.21) becomes invariant.
Under this circumstance, the diagonalized form of the spectra (3.27) is lost
as we shift the observational procedures to the primed frame, but it may be
recovered by implementing the $\bigstar $-invariant device%
\begin{equation}
A_{(D)}^{\prime (p\pm )}=\mathfrak{s}^{\pm }A^{\prime (p\pm )}\mathfrak{s}%
^{\pm \bigstar },\text{ diag }A_{\mu \nu }^{\prime (p\pm )}\doteqdot
<e_{(\mu )}^{\prime (p\pm )}\mid A_{(D)}^{\prime (p\pm )}\parallel e_{(\nu
)}^{\prime (p\pm )}>_{g^{\prime \pm }},  \tag{7.19}
\end{equation}%
which takes up the intrinsic entries $\mathfrak{s}_{\mu }^{\pm }{}^{\nu }$
of the restriction constituents of a local unitary operator $\mathfrak{s}$,
in accordance with the diagonal patterns%
\begin{equation}
\text{diag }A_{\mu \nu }^{\prime (p\pm )}=a_{\mu }^{\prime (p\pm )}g_{\mu
\nu }^{\prime \pm }\text{ (no summation over here)},  \tag{7.20}
\end{equation}%
and\footnote{%
The use of the diagonalization prescription (7.19) was alluded to in Ref.
[17].}%
\begin{equation}
\text{diag }A_{\mu \nu }^{\prime (p\pm )}=\mathfrak{s}_{\mu }^{\pm
}{}^{\lambda }A_{\lambda \sigma }^{\prime (p\pm )}\overline{\mathfrak{s}%
^{\pm \sigma }{}_{\nu }}.  \tag{7.21}
\end{equation}%
Equations (7.15) and (7.21) thus give the correlation%
\begin{equation}
\text{diag }A_{\mu \nu }^{\prime (p\pm )}=\mathfrak{s}_{\mu \lambda }^{\pm
}{}g_{\pm }^{\ast \lambda \sigma }{}u_{\sigma \rho }^{\pm }{}A_{(p\pm
)}^{\ast \rho \tau }u_{\tau \zeta }^{\pm \bigstar }g_{\pm }^{\ast \zeta \xi
}{}\mathfrak{s}_{\xi \nu }^{\pm \bigstar }{},  \tag{7.22}
\end{equation}%
which can be rapidly reset as the configuration%
\begin{equation}
\text{diag }A_{\mu \nu }^{\prime (p\pm )}=\mathfrak{s}_{\mu }^{\pm
}{}^{\lambda }u_{\lambda }^{\pm }{}^{\sigma }A_{\sigma \rho }^{(p\pm )}%
\overline{u^{\pm \rho }{}_{\tau }}\overline{\mathfrak{s}^{\pm \tau }{}_{\nu }%
},  \tag{7.23}
\end{equation}%
whence%
\begin{equation}
\det (\text{diag }A_{\mu \nu }^{\prime (p\pm )})=\det (A_{\mu \nu }^{(p\pm
)}),  \tag{7.24a}
\end{equation}%
because [17]%
\begin{equation}
\mathfrak{s}_{\mu \nu }^{\pm \bigstar }{}=g_{\mu \lambda }^{\pm }\Delta
_{\pm }^{\ast \lambda \sigma }<e_{(\sigma )}^{\prime (p\pm )}\mid \mathfrak{s%
}^{\pm \dag }\parallel e_{(\rho )}^{\prime (p\pm )}>_{g^{\prime \pm }}g_{\pm
}^{\ast \rho \tau }{}\Delta _{\tau \nu }^{\pm },  \tag{7.24b}
\end{equation}%
and%
\begin{equation}
\det (A_{\mu \nu }^{(p\pm )})=\det (A_{(p\pm )}^{\ast \mu \nu }). 
\tag{7.24c}
\end{equation}

To express the property concerning the invariance of traces, we have to
consider the adjoint prescription%
\begin{equation}
\text{diag }A_{(p\pm )}^{\ast \prime \mu \nu }=\mathfrak{S}^{\pm \mu
}{}_{\lambda }A_{(p\pm )}^{\ast \prime \lambda \sigma }\overline{\mathfrak{S}%
_{\sigma }^{\pm }{}^{\nu }},  \tag{7.25a}
\end{equation}%
with%
\begin{equation}
\overline{\mathfrak{S}_{\mu }^{\pm }{}^{\nu }}\doteqdot \mathfrak{s}_{\mu
}^{\pm \dag }{}^{\nu }.  \tag{7.25b}
\end{equation}%
Then%
\begin{equation}
\text{Tr }A^{\prime (p\pm )}=\text{Tr }A^{(p\pm )}=\text{diag }A_{\mu \nu
}^{\prime (p\pm )}\mathfrak{g}_{\pm }^{\ast \nu \mu }{},  \tag{7.26a}
\end{equation}%
and%
\begin{equation}
\text{Tr }A_{(p\pm )}^{\ast \prime }=\text{Tr }A_{(p\pm )}^{\ast }=\text{%
diag }A_{(p\pm )}^{\ast \prime \mu \nu }\mathfrak{g}_{\nu \mu }^{\pm }, 
\tag{7.26b}
\end{equation}%
where%
\begin{equation}
\mathfrak{g}_{\mu \nu }^{\pm }\doteqdot \mathfrak{s}_{\mu }^{\pm
}{}^{\lambda }g_{\lambda \sigma }^{\pm }\overline{\mathfrak{s}^{\pm \sigma
}{}_{\nu }},\text{ }\mathfrak{g}_{\pm }^{\ast \mu \nu }\doteqdot \mathfrak{S}%
^{\pm \mu }{}_{\lambda }g_{\pm }^{\ast \lambda \sigma }\overline{\mathfrak{S}%
_{\sigma }^{\pm }{}^{\nu }}.  \tag{7.26c}
\end{equation}

In any case of spectral invariance, the product (3.25) implies that
observable eigenvalues bear invariance albeit probabilities do not, whence
unstarred expectation values, say, behave like the square of absolute values
of amplitudes, according to%
\begin{equation}
<A^{\prime (p\pm )}>_{\Phi ^{\prime \pm }}=\Phi _{(p\pm )}^{\lambda
}u_{\lambda \sigma }^{\pm \bigstar }A_{(p\pm )}^{\ast \sigma \rho }u_{\rho
\tau }^{\pm }\overline{\Phi _{(p\pm )}^{\tau }}.  \tag{7.27}
\end{equation}%
This involves the correlational case of the spectra for any charges, spins,
polarizations and helicities of massless particles. The helicity spectra for
massive particles, which would circumstantially have been prepared by some
observer, will behave invariantly only when we deal with either rotations or
boost parameters that do not cause any spacetime overpasses on the rest
frames of the particles.

Equations (3.33) and (3.34) are required to be $\mathcal{P}_{Dyn}^{+\uparrow
}$-invariant, whence the action of the charge-conjugation operator (3.32)
should fulfill the matrix-entry equality of (7.16). We thus must have%
\begin{equation}
\text{$\mathbb{Q}$}^{\prime (p\pm )}=u_{Dyn}^{\pm \bigstar }\text{$\mathbb{Q}%
^{(p\pm )}$}u_{Dyn}^{\mp }.  \tag{7.28}
\end{equation}%
Evidently, the behaviours of any energy restrictions and spectra have to be
subject to (7.15) such that (3.55b) and (3.58) carry invariant
prescriptions. It may be claimed that (7.19)-(7.21) should supply the
primed-frame version of the spectral configurations (3.55c) through%
\begin{equation}
(H_{I\mu \nu }^{\prime (p\pm )})=\left( 
\begin{array}{ll}
E^{\prime } & 0 \\ 
0 & -E^{\prime }%
\end{array}%
\right) ,\text{ }(H_{II\mu \nu }^{\prime (p\pm )})=\left( 
\begin{array}{ll}
-E^{\prime } & 0 \\ 
0 & E^{\prime }%
\end{array}%
\right) .  \tag{7.29}
\end{equation}%
Accordingly, the prescription (7.21) would ensure the formal preservation of
the schemes (3.59), with $E^{\prime }\gtrless 0$ and $E^{\prime }>0$ in the
massive and massless cases, respectively.

The commutators of (3.23) possess an invariance property as (7.15) and
(7.16) lead to%
\begin{equation}
\lbrack A^{\prime (p\pm )},B^{\prime (p\pm )}]=[A^{(p\pm )},B^{(p\pm )}], 
\tag{7.30a}
\end{equation}%
and%
\begin{equation}
\lbrack A^{\prime (p\pm )},B^{\prime (p\pm )}]=u_{Dyn}^{\pm \bigstar
}{}[A^{(p\pm )},B^{(p\pm )}]u_{Dyn}^{\pm }{}.  \tag{7.30b}
\end{equation}%
If the observational characters of the commutator entries are mixed, we can
still write, for instance,%
\begin{equation}
\lbrack A^{\prime (p\pm )},B^{\prime (p\pm )}]=0\Rightarrow \lbrack A^{(p\pm
)},u_{Dyn}^{\pm \bigstar }{}B^{(p\pm )}u_{Dyn}^{\pm }{}]=0,  \tag{7.31}
\end{equation}%
which may make up the behaviour of the statement (3.22) when the operator
correlation of (7.16) applies to the $R$-restrictions. Any of the
measurements considered in Section 5 should be performed covariantly. Hence,
the matrices (5.3) must afford invariant representations while the state
reductions (5.10) can not generally exhibit invariance.

\section{Concluding remarks and outlook}

One of the most significant features of $\mathcal{P}_{Dyn}^{+\uparrow }$
makes invariant the dynamical decomposition (2.8) and the restriction
pattern (2.21). It has thereby brought together through the definition (7.1)
the underlying pseudo unitarity of $SU(2,2)$ and a restricted unitarity
property. The general $\mathcal{P}_{Dyn}^{+\uparrow }$-invariance of
degenerate spectra demonstrated explicitly in Section 7, has shown that
helicity operators for anomalous neutrinos or antineutrinos should
adequately be taken to commute with all the pseudo-unitary operators in $%
\mathfrak{C}^{\pm }$ and $\mathfrak{C}_{\pm }^{\ast }$. Any rest-mass
spectra for such fermions could have been written down by defining mass
operators that satisfy the correlation (7.16). If this procedure had been
implemented, it would not of course go against the self-commutativity
contents of Naimark's theorems. While the Pauli-Weisskopf theorem
establishes theoretically the existence of particle-antiparticle pairs of
any spin, the ordinary charge conjugations are valid only for spin one-half
particles and electric charges. This discrepancy has been overcome by the
definitions (3.31) and (3.33) which associate charge operators along with
their charge conjugations to any flavour-colour degrees of freedom.

The physically necessary condition whereby any energy spectra must be
non-degenerate could enable us to gain some fresh insights into the
discussions regarding the need for negative-energy particles even in the
uncharged massless case. We saw that negative energies have to be introduced
in order to guarantee the required non-degeneracy property without
nevertheless imparting any $PT$-character to spectral matrices or charge
conjugations. Thus, the ascription of minus signs to fermionic states, which
comes from the implementation of successive time reversals, as well as its
relationships with spin and statistics, have not occurred here.
Particularly, the energy spectra for $(p^{+},p^{-})$ were first constructed
locally with the help of the virtual-particle restrictions $v^{(p\pm )}$,
and then formally correlated to the ones for other observers through $%
\mathcal{P}_{Dyn}^{+\uparrow }$-transformations.

A noteworthy particularity of the operators $v^{(p\pm )}$ is that the
involutive correspondences $0^{\pm }1^{\pm }\leftrightarrow 1^{\pm }0^{\pm }$
invariantly supplied by them, interchange any standard up-down and
vertical-horizontal states of $p^{\pm }$ without altering the respective
spin-polarization spectra, i.e., 
\begin{equation*}
<e_{(0)}^{(p\pm )}\mid v^{(p\pm )}\Omega ^{(p\pm )}v^{(p\pm )}\parallel
e_{(0)}^{(p\pm )}>_{g^{\pm }}=<e_{(1)}^{(p\pm )}\mid \Omega ^{(p\pm
)}\parallel e_{(1)}^{(p\pm )}>_{g^{\pm }},
\end{equation*}%
and%
\begin{equation*}
<e_{(1)}^{(p\pm )}\mid v^{(p\pm )}\Omega ^{(p\pm )}v^{(p\pm )}\parallel
e_{(1)}^{(p\pm )}>_{g^{\pm }}=<e_{(0)}^{(p\pm )}\mid \Omega ^{(p\pm
)}\parallel e_{(0)}^{(p\pm )}>_{g^{\pm }},
\end{equation*}%
with $\Omega ^{(p\pm )}$ amounting to either $\Sigma ^{(p\pm )}$ or $\Pi
^{(p\pm )}$. Amongst all the considerable pseudo-Hermitian virtual-particle
matrices for $v^{(p\pm )}$, which could be formed by either intrinsic or
spectral entries, only the disjoint ones given by (3.58) simultaneously
maintain intact the set of spin-polarization values for $p^{\pm }$ and
constitute representations that do not pertain to the class of matrices
having the unimodular shape of the $\beta $-blocks of (7.2). Any attempt at
removing the pseudo Hermiticity of $v^{(p\pm )}$ from the dynamical picture
could transgress these requirements. In respect to such properties, one of
the sharpest points is that piecing together the matrices (3.58) would give
rise to an element of the $g$-realization of $SU(2,2)$, contrarily to the
overall action $0^{\pm }1^{\pm }\leftrightarrow 1^{\mp }0^{\mp }$ of $%
\mathbb{Q}^{(p^{+}p^{-})}$. Provided that any schemes like (3.58) and (3.59)
are wholly assigned to individual orthocronous-proper observers, we may
ultimately infer that every virtual-particle description must be formulated
as if particles and antiparticles were propagating alone. A similar point
should also be made for the case of the spectrum (3.54) which, like the
reduced form of $(v_{\mu \nu }^{(p\pm )})$, can not stand for any element of 
$\mathcal{P}_{Dyn}^{+\uparrow }$. If we were in principle to replace the
translational contributions $(I_{2}\pm iW)$ by $\pm iw$ with $\det w=\pm 1$,
then we could not consistently recover the pattern of the array (6.22). In
actuality, carrying out this replacement would make one unable to retrieve $%
\mathcal{L}_{+}^{\uparrow }$.

The elaboration of Section 3 has exhibited an invariant equivalence between
the frameworks of $\mathfrak{C}^{\pm }$ and $\mathfrak{C}_{\pm }^{\ast }$,
but the diagonalization treatment as prescribed by (7.21) of any spectra
conditioned by the correlation (7.15), seems to require that both
formulations should be set down conjunctively at least to some extent. Our
unitary techniques for changing locally the description of degrees of
freedom can be effectively implemented in any $\mathcal{L}_{+}^{\uparrow }$%
-frame where some copy of the adjoint computational bases for the considered
observer would have been chosen beforehand by means of a correlation like
that of (7.6). Therefore, the preservation of the configurations (3.37) and
(3.58) that may certainly be settled in by the prescription (3.63), is
ensured in the primed frame by operator associations like%
\begin{equation*}
\mathbb{Q}^{\prime (p\pm )}\mapsto \mathfrak{U}^{\pm \dag }u_{Dyn}^{\pm
\bigstar }\mathbb{Q}^{(p\pm )}u_{Dyn}^{\mp }\mathfrak{U}^{\mp },\text{ }%
v^{\prime (p\pm )}\mapsto \mathfrak{U}^{\pm \dag }u_{Dyn}^{\pm \bigstar
}v^{(p\pm )}u_{Dyn}^{\pm }\mathfrak{U}^{\pm },
\end{equation*}%
which take into account the behavioural law (7.28) and likewise retain the
property (3.47).

As emphasized in Ref. [15], a striking feature of the conventional particle
theories brings out the fact that whilst the operator description of spin
one-half particles usually precedes the achievement of the specification of
the corresponding spin states, the classical description of the possible
polarization states for photons is what normally carries an immediate
physical meaning. Such a contextual contraposition has not taken place in
Sections 3 and 4 as the fermionic and bosonic procedures allowed for there
were carried out on the same footing. Indeed, the entire construction of
Sections 6 and 7 has not involved the utilization of any of the generators
of $\mathcal{P}_{+}^{\uparrow }$. The procedure that assigns copies of $%
\mathfrak{C}^{\pm }$ and $\mathfrak{C}_{\pm }^{\ast }$ to any particles or
antiparticles resembles the one mentioned in Section 1 which uses unitary
irreducible representations of $\mathcal{P}_{+}^{\uparrow }$ for
constructing observational correlations in flat spacetime for any free
quantum mechanical systems. By just taking up the maximal spacetime symmetry
ascribed to $\mathcal{C}_{+}^{\uparrow }$, we have rectified the situation
related to the earlier theoretical absence of geometric decompositions that
might account for a combined version of the dynamics of free particles and
antiparticles. Free fermions and bosons can thus be described covariantly
together with their antiparticles within the same symbolic framework, in
contradistinction with the traditional quantum mechanical contexts. This
unifying characteristic may be useful for phenomenological purposes since it
affords the possibility of comparing easily formal conservation laws for
scattering processes with available experimental data, and evaluating
entropies of Feynman diagrams in a systematic manner. Hence, a definition of
mixture of composite states that extends the non-relativistic one may be
used to manipulate states for channels of particle reactions in any $%
\mathcal{L}_{+}^{\uparrow }$-frame. It is evident that the description of
quarks, gluons and electroweak bosons becomes physically accomplishable
before the occurrence of hadronizations. In typical cases, the observational
correlations between the descriptions of scattering processes could take up
some boost-translation constituents of $\mathcal{P}_{+}^{\uparrow }$ while
the rotation-translation choice could be made when the implementation of the
dynamical-group action follows some local preparations of states.

All the methodological statements we have derived previously repose
principally upon the claim that the most natural quantum mechanical
framework for free elementary particles which may be conceived at present
should emerge from the combination of the twofold pseudo-unitary structures
provided by special relativity with the disturbance property of measurement
processes and a generalized Born rule for composite amplitudes. The
customary interpretations of Stern-Gerlach and photon-detection experiments
should accordingly be taken to bear a universal character. Since the
beginning of the development of the programme considered in Section 1, many
works based on our approach and devoted to the description of quantum
computational processes have been sketched out. We could find it very much
interesting, in particular, to implement this approach for drawing up
covariant computational gates and quantum circuits. We think that the
availability of the procedures for handling the $\mathcal{P}%
_{Dyn}^{+\uparrow }$-behaviours of the amplitudes of suitably prepared
entangled states, should motivate a careful and necessary revision of the
existing expressions concerning the quantum-theoretical locality and
non-locality issues. As we believe, such investigations may bring forth a
clear concept of antientropy in a fresh relativistic domain of quantum
information theory. These situations will perhaps be entertained elsewhere.

\end{document}